\documentclass[lettersize,journal]{IEEEtran}
\IEEEoverridecommandlockouts
\usepackage{makecell}
\usepackage{array}
\usepackage{graphicx,amssymb,amsmath}
\usepackage{multicol}
\usepackage[noadjust]{cite}
\usepackage{setspace}
\usepackage{subfigure}
\usepackage{graphicx}
\usepackage{float}
\usepackage{url}
\usepackage{stfloats}
\usepackage{amsthm,pifont}
\usepackage{flushend}
\usepackage{cases,subeqnarray}
\usepackage{bm,multirow,bigstrut}
\usepackage{amsmath, amsthm, amssymb}
\usepackage{textcomp}
\usepackage{latexsym,bm}
\usepackage{booktabs}
\usepackage{xcolor}
\usepackage{mathtools}
\usepackage{dsfont}
\usepackage{extarrows}
\usepackage{epsfig}
\usepackage{epstopdf}
\usepackage[noend]{algpseudocode}
\usepackage{algorithmicx}

\usepackage{algorithm}
\usepackage{algpseudocode}
\usepackage{amsmath}
\usepackage{pifont}
\usepackage{cite}
\usepackage{bm}
\usepackage{cleveref}
\usepackage{multicol}       
\usepackage{multirow}       
\usepackage{array}          
\usepackage{colortbl}
\usepackage{makecell}
\definecolor{crimson}{RGB}{192,0,0}         
\definecolor{navy}{RGB}{47,85,151}         
\usepackage{bbding}
\usepackage{graphicx}
\usepackage{booktabs}
\usepackage{algorithm}
\usepackage{algpseudocode}

\theoremstyle{plain}

\theoremstyle{plain}
\newtheorem{rem}{Remark}

\usepackage{amsmath}

\IEEEoverridecommandlockouts
\begin{document}

\title{Low-Complexity Multi-Agent Continual Learning for Stacked Intelligent Metasurface-Assisted Secure Communications}

\author{Enyu Shi*, Yiyang Zhu*, Jiayi~Zhang,~\IEEEmembership{Senior Member,~IEEE}, Ziheng Liu, Jiakang Zheng, Jiancheng An,~\IEEEmembership{Member,~IEEE}, Derrick Wing Kwan Ng,~\IEEEmembership{Fellow,~IEEE}, Bo Ai,~\IEEEmembership{Fellow,~IEEE}, and Chau Yuen,~\IEEEmembership{Fellow,~IEEE}
\thanks{*Enyu Shi and Yiyang Zhu contributed equally to this work.}
\thanks{E. Shi, J. Zhang, Z. Liu, J. Zheng, and B. Ai are with the State Key Laboratory of Advanced Rail Autonomous Operation, and also with the School of Electronics and Information Engineering, Beijing Jiaotong University, Beijing 100044, P. R. China, and also with Nanjing Rongcai Transportation Technology Research Institute Co., Ltd., Nanjing 210000, China. (e-mail: \{enyushi, jiayizhang, zihengliu, jiakangzheng, boai\}@bjtu.edu.cn).}
\thanks{D. W. K. Ng is with the School of Electrical Engineering and Telecommunications, University of New South Wales, NSW 2052, Australia. (e-mail: w.k.ng@unsw.edu.au).}
\thanks{Y. Zhu, J. An and C. Yuen are with the School of Electrical and Electronics Engineering, Nanyang Technological University, Singapore 639798 (e-mail: yiyang015@e.ntu.edu.sg, jiancheng\_an@163.com, chau.yuen@ntu.edu.sg).}
}
\maketitle
\begin{abstract}
Stacked intelligent metasurfaces (SIMs), composed of multiple layers of reconfigurable transmissive metasurfaces, are gaining prominence as a transformative technology for future wireless communication security. This paper investigates the integration of SIM into multi-user multiple-input multiple-output (MIMO) systems to enhance physical layer security. A novel system architecture is proposed, wherein each base station (BS) antenna transmits a dedicated single-user stream, while a multi-layer SIM executes wave-based beamforming in the electromagnetic domain, thereby avoiding the need for complex baseband digital precoding and significantly reducing hardware overhead. To maximize the weighted sum secrecy rate (WSSR), we formulate a joint precoding optimization problem over BS power allocation and SIM phase shifts, which is high-dimensional and non-convex due to the complexity of the objective function and the coupling among optimization variables. To address this, we propose a manifold-enhanced heterogeneous multi-agent continual learning (MHACL) framework that incorporates gradient representation and dual-scale policy optimization to achieve robust performance in dynamic environments with high demands for secure communication. Furthermore, we develop SIM-MHACL (SIMHACL), a low-complexity learning template that embeds phase coordination into a product manifold structure, reducing the exponential search space to linear complexity while maintaining physical feasibility. Simulation results validate that the proposed framework achieves millisecond-level per-iteratio ntraining in SIM-assisted systems, significantly outperforming various baseline schemes, with SIMHACL achieving comparable WSSR to MHACL while reducing computation time by 30\%.
\end{abstract}


\IEEEpeerreviewmaketitle

\section{Introduction}\label{introduction}

\IEEEPARstart{T}{he} upcoming sixth-generation (6G) networks are anticipated to play a pivotal role in shaping future societies, industries, and daily life, demanding exceptionally high standards for network capacity, latency, reliability, and intelligence \cite{wang2023road}. Despite the considerable convenience and numerous benefits provided by wireless communications, significant challenges related to information security have emerged that can no longer be overlooked \cite{wang2024active}. In particular, secure and covert transmission technologies have become indispensable with the rapid development of the Internet of Things (IoT), where transmitted data frequently includes highly sensitive personal information, such as medical and health records \cite{sun2024secure}. Within the boarder scope of secure communication, physical layer security (PLS) has garnered substantial research interest. In fact, numerous advanced techniques, including distributed multiple-input multiple-output (MIMO) \cite{xing2023joint}, extra-large-scale/holographic MIMO \cite{wang2024tutorial}, and reconfigurable intelligent surfaces (RIS) \cite{10556753}, have been recently proposed to strengthen security at the physical layer, ensuring robust protection while simultaneously delivering exceptional performance, e.g., achieving sufficiently high secrecy rates.

Among these technologies, RISs have demonstrated substential potential for enhancing secure communications \cite{wei2024star,wu2019intelligent}. Specifically, with a large number of low-cost nearly-passive elements, a RIS is able to reflect incident electromagnetic (EM) signals toward desired directions, providing high array gains by adaptive phase shift across its reflecting elements according to channel conditions \cite{10679239}. For example, the authors in \cite{li2025covert} proposed a dual-function RIS architecture that supports cooperative relaying for improving secure communications. In particular, the results highlight that increasing the number of elements can substantially improve security performance and effectively mitigate the detrimental impact of RIS phase errors. Besides, in \cite{xiao2023simultaneously}, the authors investigated a multi-antenna covert communications assisted by a simultaneously transmitting and reflecting RIS (STAR-RIS), where a secure rate maximization problem was formulated considering quality of service (QoS) constraints on both security and communication, incorporating outage probability analysis. However, the authors in \cite{hao2022securing} studied PLS in RIS-aided cell-free networks, formulating a maximum weighted sum secrecy rate problem by jointly optimizing the active beamforming at the base stations (BSs) and passive beamforming at RISs. These studies collectively indicate that optimized phase shifts in RISs are effective in improving the quality of PLS-based communications. In fact, recent system-level simulation results have further showcased the performance gains provided by RIS in practical cellular network deployments, provided that the surface size is sufficiently large \cite{sihlbom2022reconfigurable}. Typically, RIS nodes are deployed distributly across the environment, motivating various ongoing research efforts for reducing channel estimation and control overhead of RIS-assisted communications \cite{9244106}. Despite these significant performance improvements, RIS technology faces inherent limitations. As a low-complexity and nearly passive ``transparent" electronic component, a RIS cannot support active transmit/receive functionalities, and thus lacks the spatial multiplexing capabilities of conventional active multi-antenna base stations (BSs) \cite{an2021low}.

Motivated by recent advancements in metasurface technology, researchers have progressed beyond traditional RIS implementations and started investigating multi-layer intelligent surfaces for EM wave-based signal processing \cite{doi:10.1126/science.aat8084,liu2022programmable,nerini2024physically,liu2024stacked,shi2025downlink}. Specifically, these novel structures function as diffractive media, facilitating various applications such as low-complexity and energy-efficient MIMO transceiver architectures.
For instance, in \cite{liu2022programmable}, the authors proposed a programmable deep neural network structure comprising multiple metasurface layers, where each individual meta-atom serves as an adaptable artificial neuron. Furthermore, extending this approach, researchers in \cite{10158690} developed a novel stacked intelligent metasurface (SIM)-based MIMO transceiver architecture, which employs multiple programmable, nearly-passive metasurface layers to implement wireless transceivers that mimic the behavior of artificial neural networks (ANNs). Indeed, this innovative SIM technology enables advanced wave-based signal processing at the speed of light, effectively performing data encoding and decoding operations in the EM domain.Specifically, by adaptively configuring paired SIM modules, complex multi-stream power allocation processes can be transformed from conventional digital implementations into simplified analog operations conducted entirely in the wave domain. Consequently, this approach markedly reduces both computational complexity and power consumption compared to traditional fully-digital transmitters.
In addition, the authors in \cite{nadeem2023hybrid} utilized SIM-based designs for efficient receiver combining and transmit power allocation in holographic MIMO systems, demonstrating that multi-layer metasurface architectures offer clear advantages over traditional single-layer counterparts in terms of performance and flexibility.
These studies demonstrate that SIM-based transceiver architectures can effectively emulate the performance gains typically associated with increasing the number of transceiver antennas, while simultaneously supporting efficient power allocation \cite{li2025sim, wang2024multi}. This paradigm not only enhances available spatial degrees of freedom but also subsequently lowers power consumption, aligning well with the design goals of energy-efficient and sustainable wireless communication systems. 

With the growing exploration of SIM-assisted next-generation communication systems, research on leveraging SIM for enhancing PLS has recently gained increasing interests \cite{liu2025stacked}. For example, the authors in \cite{pei2024stacked} proposed a SIM-assisted integrated sensing and resistance (ISAR) anti-jamming framework. Within this scheme, a transmissive RIS (TRIS)-based optimization algorithm was introduced, which dynamically adapts the phase shifts based on the acquired jamming information, thereby enabling effective interference signals. Besides, the authors in \cite{niuab2024enhancing,niu2024efficient} investigated the performance of a SIM-assisted single-input single-output (SISO) system by designing the joint precoding strategies. 
The above studies preliminarily demonstrate that SIM offers significant potential to strengthening PLS. However, most existing research remain limited on single-user scenarios, while in practice, communication typically involves multiple users. In such cases, the security challenges become more complex, thereby placing higher demands on the joint design of the SIM phase shifts and BS precoding. 

Fortunately, artificial intelligence (AI)-driven optimization frameworks, particularly deep reinforcement learning (DRL), have demonstrated remarkable promise for facilitating secure communication provisioning through their inherent capability to handle high-dimensional non-convex optimization problems in dynamic environments \cite{yu2024nature,alexandropoulos2022pervasive}. This unique characteristic aligns perfectly with the multi-layer architecture of SIMs, as the joint optimization of discrete phase shifts across cascaded metasurface layers establishes an exponentially large search solution space that conventional optimization methods struggle to efficiently navigate  \cite{zhang2025marl,zhu2024robust}. Meanwhile recent studies have validated learning-based algorithms' potential in physical layer security. For instance, the authors in \cite{dong2024secure} proposed a DRL-based framework for RIS-assisted unmanned aerial vehicle (UAV) networks that achieves a 63\% secrecy rate improvement through intelligent phase shift coordination and trajectory optimization. Fortunately, in multi-UAV scenarios, \cite{guo2024ris} developed a hybrid DRL approach that simultaneously optimizes RIS configurations and artificial noise injection, demonstrating a 42\% reduction in eavesdropping probability compared with conventional optimization methods. Additionally, the survey in \cite{khoshafa2024ris} further confirmed that machine learning techniques can effectively address the dynamic security challenges inherent in RIS-assisted systems through real-time environment adaptation. These advancements highlight that AI-driven methods provide a viable and effective pathway to address the joint optimization challenges of secure beamforming and environment-aware configuration in SIM-enabled systems. However, existing learning-based approaches encounter three critical limitations in SIM-assisted secure communications: (i) inability to effectively handle the coupled optimization of discrete phase shifts and continuous power allocation, (ii) lack of continual adaptation mechanisms for non-stationary wireless environments, and (iii) computational complexity that grows exponentially with the number of SIM metasurface layers.

Motivated by the aforementioned observations, we develop a comprehensive modeling framework and conduct a thorough analysis of the secrecy rate of SIM-assisted multi-user MIMO systems. The key contributions of this work are delineated as follows:

\begin{itemize}
\item We introduce a novel integration of SIM into multi-user MIMO communication systems, to further enhance their secrecy rate. In the proposed system, a multi-layer SIM capable of performing signal processing in the EM domain is equipped at the BS, enabling each antenna to independently transmit a dedicated data stream to an individual user. By combining wave-based beamforming with direct power control, this approach eliminates the reliance on complex baseband digital precoding, significantly reducing the BS's hardware and computational complexity. Based on this architecture, a joint BS antenna power control and SIM beam-domain precoding design problem is formulated to maximize the weighted sum secrecy rate (WSSR). 

\item To address the WSSR problem, a manifold-enhanced heterogeneous multi-agent continual learning (MHACL) framework is proposed, which can effectively address environmental non-stationarity, integrating dynamic gradient masking and prioritized experience buffers to enable secure and adaptive optimization. Within this framework, channel state information (CSI) is represented as gradient tensors in the state space, inherently obfuscating sensitive channel details while preserving optimization dynamics. Furthermore, a dual-scale policy architecture alternates between power allocation and phase shift optimization via Riemannian manifold projections, thereby enforcing unit-modulus constraints while decoupling coupled variables, ensuring adaptive secrecy rate maximization under strict unit-modulus constraints.  

\item Furthermore, we propose SIM-MHACL (SIMHACL), a low-complexity algorithmic template derived from the MHACL framework, embedding the multi-layer SIM phase shifts into a product manifold structure. By reformulating the exponential search solution space of conventional alternating optimization into a linearly scalable manifold projection, the proposed approach guarantees convergence while simultaneously preserving physical feasibility, eliminating explicit quantization steps, and aligning with wave-based beamforming principles.

\item Simulation results demonstrate that the integration of SIM can significantly improve the system secrecy rate. In particular, the proposed MHACL framework achieves millisecond-level training time per iteration in SIM-assisted systems. Furthermore, the low-complexity SIMHACL algorithm achieves an approximately 30\% reduction in the computational complexity time at the cost of only a slight performance degradation.

\end{itemize}

The remainder of this paper is structured as follows. Section \uppercase\expandafter{\romannumeral2} outlines the detailed SIM-assisted mult-user MIMO system model, including the channel model, downlink data transmission, and the corresponding problem formulation. Next, Section \uppercase\expandafter{\romannumeral3} proposes the learning-based optimization framework, consisting of manifold theory and heterogeneous agent continual learning. Then, numerical results are illustrated and discussed in Section \uppercase\expandafter{\romannumeral4}. Finally, Section \uppercase\expandafter{\romannumeral5} concludes this paper.

\textbf{Notation:} Column vectors and matrices are denoted by boldface lowercase letters $\mathbf{x}$ and boldface uppercase letters $\mathbf{X}$, respectively. The superscripts $\mathbf{x}^{\rm{H}}$, $\mathbf{x}^\mathrm{T}$, and $\mathbf{x}^\mathrm{*}$ represent the conjugate transpose, transpose, and conjugate, respectively. The symbols $\triangleq$, $\left\|  \cdot  \right\|$, and $\left\lfloor  \cdot  \right\rfloor $ denote the definitions, the Euclidean norm, and the floor function, respectively. ${\rm{tr}}\left(  \cdot  \right)$, $\mathbb{E}\left\{  \cdot  \right\}$, and ${\rm{Cov}}\left\{  \cdot  \right\}$ denote the trace, expectation, and covariance operators, respectively. ${\text{diag}}\left( {{a_1}, \cdots ,{a_n}} \right)$ denotes a diagonal matrix with diagonal entries ${{a_1}, \cdots ,{a_n}}$. ${\left[ a \right]^ + } = \max \left( {a,0} \right)$. A circularly symmetric complex Gaussian random variable $x$ with mean $0$ and variance $\sigma^2$ is denoted by $x \sim \mathcal{C}\mathcal{N}\left( {0,{\sigma^2}} \right)$. $\nabla$ denotes the gradient operator and $\mathcal{O}(\cdot)$ denotes the big-O notation, respectively. $\mathbb{B}^n$, $\mathbb{Z}^n$, $\mathbb{R}^n$, and $\mathbb{C}^n$ represent the $n$-dimensional spaces of binary, integer, real, and complex numbers, respectively. Finally, the $N \times N$ zero matrix and identity matrix are denoted by $\mathbf{0}_{N}$ and $\mathbf{I}_{N}$, respectively.


\begin{figure}[t]
\centering
\includegraphics[scale=1]{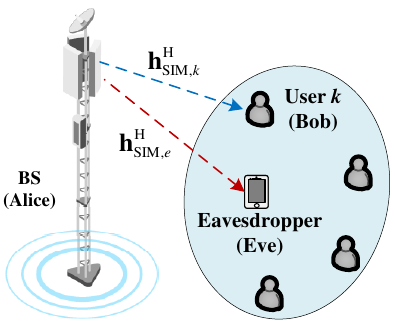}
\caption{Illustration of SIM-assisted multi-user secure communication systems.} \label{system_model}
\end{figure}

\section{System Model}\label{se:model}
As shown in Fig.~\ref{system_model}, we consider a multiuser MIMO secure communication system that consists of a SIM-assisted legitimate transmitter (Alice), $K$ legitimate users (Bobs), and an eavesdropper (Eve), which attempts to eavesdrop on information from Alice. Alice is equipped with $L$ antennas, while Bob and Eve are single antenna devices. Specifically, we assume that the SIM is equipped with $M$ metasurface layers and $N$ meta-atoms in each layer. Let ${\cal L} = \{ 1, \ldots ,L\}$, ${\cal K} = \{ 1, \ldots ,K\}$, ${\cal M} = \{ 1, \ldots ,M\}$, and ${\cal N} = \{ 1, \ldots ,N\}$ denote the index sets of Alice antennas, Bobs, SIM metasurface layers, and meta-atoms per layer, respectively. Furthermore, the SIM is connected to an intelligent controller at Alice, capable of applying a distinct and adaptive phase shift to the EM waves passing through each meta-atom \cite{an2023stacked2}.

\subsection{Channel Model}\label{Channel}
We first explain the working principle of the SIM-assisted legitimate transmitter. As shown in Fig.~\ref{system_model}, with the deployment of a SIM, baseband digital precoding is no longer required at the BS \cite{an2023stacked2}. Instead, only low-resolution digital-to-analog converters (DACs) are needed, significantly reducing the hardware costs of the BS. Specifically, each data stream is transmitted through a dedicated antenna after passing through the low-resolution DACs and RF chains, subsequently impinging upon the first layer of the SIM. In accordance with the Huygens–Fresnel principle \cite{doi:10.1126/science.aat8084}, the EM wave passing through each meta-atom in a given layer acts as a secondary point source, illuminating all meta-atoms in the subsequent layer. Furthermore, the EM waves impinging on a meta-atom in a metasurface layer are superimposed, acting as a wave incident onto this meta-atom. Finally, the EM wave exits the last metasurface layer and propagates to the users through the wireless channel.

\begin{figure}[t]
\centering
\includegraphics[scale=0.8]{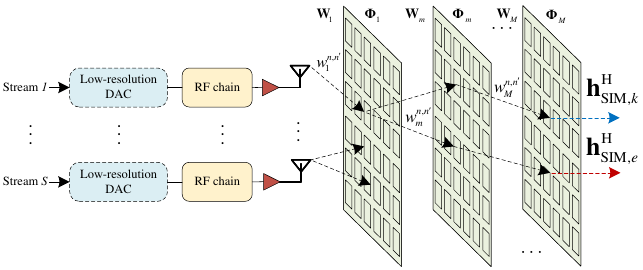}
\caption{Illustration of hardware structure and signal propagation process of SIM-assisted BS.}\label{signal_transmission} 
\end{figure}

Due to the fact that a SIM is a device composed of a series of low-cost and low-power metamaterial electromagnetic meta-atoms, it is often difficult to achieve an ideal continuous phase in hardware. Therefore, we consider a discrete phase-shift model for the SIM. 
Specifically, we adopt ${e^{j\varphi _{m}^n}},\forall m \in {\cal M},\forall n \in {\cal N}$, with
\begin{align}
    \varphi _m^n = \frac{{i_m^n}}{{{2^{b}}}}\pi ,i_m^n \in \left\{ {0,1, \ldots ,{2^b} - 1} \right\},
\end{align}
to denote the $n$-th meta-atom's phase shift in the $m$-th metasurface layer, where $b$ denotes the number of quantization bits. Hence, the diagonal phase shift matrix ${{\bf{\Phi }}_{m}}$ for the $m$-th metasurface layer at the SIM can be denoted as ${{\bf{\Phi }}_{m}} = {\rm{diag}}( {{e^{j\varphi _{m}^1}},{e^{j\varphi _{m}^2}}, \ldots ,{e^{j\varphi _{m}^N}}} ) \in \mathbb{C} {^{N \times N}},m \in {\cal M}$. Furthermore, let ${\bf{W}}_{1}^{} = {[ {{\bf{w}}_{1}^1, \ldots ,{\bf{w}}_{1}^{L}} ]} \in \mathbb{C} {^{N \times L}}$ denote the propagation coefficients from the Alice antenna to the first metasurface layer of the SIM, where ${{\bf{w}}_{1}^{{l}}} \in \mathbb{C} {^{N}}$ denotes the propagation coefficient matrix from the $l$-th antenna of Alice to the first metasurface layer within the SIM \cite{11182313}. Let ${{\bf{W}}_{m}} \in \mathbb{C} {^{N \times N}},\forall m \ne 1,m \in {\cal M}$ denote the propagation coefficient matrix from the $(m-1)$-th to the $m$-th metasurface layer of SIM. 
As shown in Fig.~\ref{signal_transmission}, according to the Rayleigh-Sommerfeld diffraction theory introduced by \cite{doi:10.1126/science.aat8084} and \cite{10158690}, the $\left( {n,n'} \right)$-th element of matrix $\mathbf{W}_{m}$ is expressed as
\begin{align}\label{w}
w_{m}^{n,n'} = \frac{{{d_x}{d_y}\cos \chi _{m}^{n,n'}}}{{d_{m}^{n,n'}}}\left( {\frac{1}{{2\pi d_{m}^{n,n'}}} - j\frac{1}{\lambda }} \right){e^{j2\pi \frac{{d_{m}^{n,n'}}}{\lambda }}},
\end{align}
where $\lambda$ represents the carrier wavelength, $d_{m}^{n,n'}$ indicates the corresponding transmission distance, $d_{x} \times d_{y}$ indicates the size of each SIM meta-atom, and ${\chi _{m}^{n,n'}}$ represents the angle between the propagation direction and the normal direction of the $(m-1)$-th  metasurface layer of SIM. Similarly, the $n$-th element $w_{1,n}^l$ of ${{\bf{w}}_{1}^{{l}}}$ can be obtained from \eqref{w}.
Hence, the wave-based beamforming matrix ${{\bf{G}}} \in \mathbb{C}{^{N \times N}}$ of Alice, enabled by SIM, can be formulated as \cite{an2024stacked}
\begin{align}\label{G}
{{\bf{G}}} = {{\bf{\Phi }}_{M}}{{\bf{W}}_{M}}{{\bf{\Phi }}_{M - 1}}{{\bf{W}}_{M - 1}} \ldots {{\bf{\Phi }}_{2}}{{\bf{W}}_{2}}{{\bf{\Phi }}_{1}}.
\end{align}

Next, we consider a quasi-static flat-fading channel model. 
Let ${\bf{h}}_{{\rm{SI}}{{\rm{M}}},k} \in \mathbb{C}{^{N \times 1}}$ and ${\bf{h}}_{{\rm{SI}}{{\rm{M}}},e} \in \mathbb{C}{^{N \times 1}}$ denote the direct channels between the last SIM metasurface layer to Bob $k$ and Eve, respectively. Specifically, we assume that ${\bf{h}}_{{\rm{SI}}{{\rm{M}}},k}$ and ${\bf{h}}_{{\rm{SI}}{{\rm{M}}},e}$ are characterized as the spatially correlated Rayleigh fading channel with ${\bf{h}}_{{\rm{SI}}{{\rm{M}}},k} \sim {\cal C}{\cal N}\left( {\bf{0},{{\bf{R}}_{\rm{SIM},k}}} \right)$ and ${\bf{h}}_{{\rm{SI}}{{\rm{M}}},e} \sim {\cal C}{\cal N}\left( {\bf{0},{{\bf{R}}_{{\rm{SI}}{{\rm{M}}},e}}} \right)$. ${{\bf{R}}_{{\rm{SI}}{{\rm{M}}},k}} = {\beta _{{\rm{SI}}{{\rm{M}}},k}}{\bf{R}} \in \mathbb{C}{^{N \times N}}$ and ${{\bf{R}}_{{\rm{SI}}{{\rm{M}}},e}} = {\beta _{{\rm{SI}}{{\rm{M}}},e}}{\bf{R}} \in \mathbb{C}{^{N \times N}}$ where ${\beta _{{\rm{SI}}{{\rm{M}}},k}}$ denotes the distance-dependent path loss between Bob $k$ and the SIM, ${\beta _{{\rm{SI}}{{\rm{M}}},e}}$ denotes the distance-dependent path loss between Eve and the SIM, and the covariance matrix ${\bf{R}} \in \mathbb{C}{^ {N \times N}}$ describes the spatial correlation among the meta-atoms of the final metasurface layer within the SIM. Considering an isotropic scattering environment with multipath components uniformly distributed, the $\left( {n,n'} \right)$-th element of $\mathbf{R}$ is ${{\bf{R}}_{n,n'}} = {\rm{sinc}}\left( {2{{{d_{n,n'}}} \mathord{\left/
 {\vphantom {{{d_{n,n'}}} \lambda }} \right.
 \kern-\nulldelimiterspace} \lambda }} \right)$ \cite{bjornson2020rayleigh}, where $d_{n,n'}$ denotes the spacing distance between the meta-atoms and ${\rm{sinc}}\left( x \right) = {{\sin \left( {\pi x} \right)} \mathord{\left/
 {\vphantom {{\sin \left( {\pi x} \right)} {\left( {\pi x} \right)}}} \right.
 \kern-\nulldelimiterspace} {\left( {\pi x} \right)}}$ denotes the normalized sinc function. Then, the composite channel ${{\bf{h}}_{k}} \in \mathbb{C}{^{L \times 1}}$ between Alice and Bob $k$ and composite channel ${{\bf{h}}_{e}} \in \mathbb{C}{^{L \times 1}}$ between Alice and Eve can be respectively denoted as
\begin{align}\label{h_k}
{{\bf{h}}_{k}} &= {\bf{W}}_{1}^{\rm{H}}{\bf{G}}^{\rm{H}}{\bf{h}}_{{\rm{SI}}{{\rm{M}}},k},
\end{align}
\begin{align}\label{h_e}
{{\bf{h}}_{e}} &= {\bf{W}}_{1}^{\rm{H}}{\bf{G}}^{\rm{H}}{\bf{h}}_{{\rm{SI}}{{\rm{M}}},e}.
\end{align}
\begin{rem}
    We observe from \eqref{h_k} and \eqref{h_e} that the SIM effectively alters the channel state by dynamically adjusting the phase shifts within $\bf G$. In contrast to conventional MIMO systems, where the direct channels are uncontrollable, the SIM channel benefits from controllable signal aggregation across its multiple metasurface layers. This capability enables the mitigation of inter-user interference and eavesdropping prevention as EM waves traverse the SIM structure. Additionally, in the considered downlink scenario, signals reach users through the meta-atoms of the SIM's final layer. Consequently, the number of meta-atoms in this last layer effectively determines the number of transmit antennas, significantly enhancing the transmitter's spatial degrees of freedom and beamforming performance.
\end{rem}

\subsection{Downlink Data Transmission}\label{DL_data}
In contrast to conventional digital power allocation, where each transmitted symbol is typically mapped onto a dedicated beamforming vector, we adopt a wave-based beamforming approach utilizing the SIM. Under this scheme, data streams are directly transmitted from the Alice antennas. Typically, the number of Alice's antennas $L$ exceeds the number of Bobs $K$, necessitating an antenna selection process beforehand~\cite{sanayei2004antenna}. For cognitive simplicity, we assume the scenario $L = K$, where each antenna independently modulates and transmits a single data stream. Based on this assumption, the transmission matrix from the Alice antenna to the SIM first layer board can be rewritten as ${\bf{W}}_{1}^{} = {[ {{\bf{w}}_{1}^1, \ldots ,{\bf{w}}_{1}^{K}} ]} \in \mathbb{C} {^{N \times K}}$. Consequently, employing low-resolution analog-to-digital converters (ADCs) and digital-to-analog converters (DACs) becomes practical, as it offers reduced hardware complexity with only minor performance degradation.

Without loss of generality, the information symbol intended for the $k$-th Bob is denoted by ${s_k} \sim \mathcal{C}\mathcal{N}\left( {0,{p_k}} \right),\forall k \in \mathcal{K}$, which is an independent and identically distributed (i.i.d.) random variable with zero mean and unit variance. Here, $p_k$ denotes the transmit power allocated to the $k$-th Bob. Then, the total transmit power constraint at Alice reads as $\sum\nolimits_{k = 1}^K {{p_k}}  \leqslant {P_A}$, where $P_A$ is the transmit power budget of the Alice. Furthermore, by superimposing the signals that propagate through the SIM, the composite signal $y_k$ received at the $k$-th Bob is expressed as 
\begin{align}
  {y_k} &= {\mathbf{h}}_{{\text{SIM}},k}^{\text{H}}{\mathbf{G}}\sum\limits_{k' = 1}^K {{\mathbf{w}}_1^{k'}} \sqrt {{p_{k'}}} {s_{k'}} + {n_k} \notag \\
   &= {\mathbf{h}}_{{\text{SIM}},k}^{\text{H}}{\mathbf{Gw}}_1^k\sqrt {{p_k}} {s_k} + {\mathbf{h}}_{{\text{SIM}},k}^{\text{H}}{\mathbf{G}}\!\!\!\!\!\!\sum\limits_{k' = 1,k' \ne k}^K \!\!\!\!\!\!{{\mathbf{w}}_1^{k'}} \sqrt {{p_{k'}}} {s_{k'}} + {n_k}, 
\end{align}
where ${n_k} \sim \mathcal{C}\mathcal{N}\left( {0,\sigma _k^2} \right)$ denotes the additive white Gaussian noise, where ${\sigma _k^2}$ denotes the receiver noise power at the $k$-th Bob.
At the same time, the $k$-th Bob’s information wiretapped by the Eve can be written as 
\begin{align}
  y_k^e &= {\mathbf{h}}_{{\text{SIM}},e}^{\text{H}}{\mathbf{G}}\sum\limits_{k' = 1}^K {{\mathbf{w}}_1^{k'}} \sqrt {{p_{k'}}} {s_{k'}} + {n_e}, \notag \\
   &= {\mathbf{h}}_{{\text{SIM}},e}^{\text{H}}{\mathbf{Gw}}_1^k\sqrt {{p_k}} {s_k} + {\mathbf{h}}_{{\text{SIM}},e}^{\text{H}}{\mathbf{G}}\!\!\!\!\!\!\sum\limits_{k' = 1,k' \ne k}^K \!\!\!\!\!\!{{\mathbf{w}}_1^{k'}} \sqrt {{p_{k'}}} {s_{k'}} + {n_e}, 
\end{align}
where ${n_e} \sim \mathcal{C}\mathcal{N}\left( {0,\sigma _e^2} \right)$ denotes the additive white Gaussian noise, where ${\sigma _e^2}$ denotes the receiver noise power at the Eve. 
Therefore, the signal-to-interference-plus-noise-ratio (SINR) of the received signal of the $k$-th Bob and Eve eavesdropping on the information of Bob $k$ can be calculated as \cite{ng2014robust}
\begin{align}\label{SINR_k}
    {\gamma _k} = \frac{{{p_k}{{\left| {{\mathbf{h}}_{{\text{SIM}},k}^{\text{H}}{\mathbf{Gw}}_1^k} \right|}^2}}}{{\sum\nolimits_{k' \ne k}^K {{p_{k'}}{{\left| {{\mathbf{h}}_{{\text{SIM}},k}^{\text{H}}{\mathbf{Gw}}_1^{k'}} \right|}^2} + \sigma _k^2} }},\forall k \in \mathcal{K},
\end{align}
\begin{align}\label{SINR_e}
    \gamma _k^e = \frac{{{p_k}{{\left| {{\mathbf{h}}_{{\text{SIM}},e}^{\text{H}}{\mathbf{Gw}}_1^k} \right|}^2}}}{{\sum\nolimits_{k' \ne k}^K {{p_{k'}}{{\left| {{\mathbf{h}}_{{\text{SIM}},e}^{\text{H}}{\mathbf{Gw}}_1^{k'}} \right|}^2} + \sigma _e^2} }},\forall k \in \mathcal{K}.
\end{align}
Note that since each antenna transmits a single user's data stream, only power control is required. The phase and amplitude of the single data stream are adjusted through each layer of the SIM metasurface to achieve effective wave-based beamforming. 

\subsection{Problem Fomulation}\label{Problem}
In general, the design objective is to maximize the secrecy rate of the users by effectively balance the received signal strength at the intended users, and the SINR at the Eve. Therefore, based on \cite{hao2022securing}, the achievable secrecy rate for the transmission from Alice to the $k$-th Bob can be expressed as
\begin{align}
    {R_{c,k}} = {\left[ {{{\log }_2}\left( {1 + {\gamma _k}} \right) - {{\log }_2}\left( {1 + \gamma _k^e} \right)} \right]^ + }.
\end{align}
In this paper, we aim to maximize the weighted sum secrecy rate (WSSR) by jointly optimizing the power control coefficients $p_k$ and the SIM wave-based beamforming matrix ${\bf{\Phi}}_{m}$, while considering user communication service quality and eavesdropping rate limitations. Mathematically, the optimization problem can be formulated as follows
\begin{subequations}\label{P_0}
  \begin{align}
  &{\mathcal{P}^0}:\mathop {\max }\limits_{{p_k},{{\mathbf{\Phi }}_m}} \,{R_c} = \sum\limits_{k = 1}^K {{\eta _k}} \left( {{{\log }_2}\left( {1 + {\gamma _k}} \right) - {{\log }_2}\left( {1 + \gamma _k^e} \right)} \right) \hfill \label{0-1}\\
  &\quad \quad\quad {\text{s}}{\text{.t}}{\text{.}}\;\;\sum\nolimits_{k = 1}^K {{p_k}}  \leqslant {P_A},  \label{0-2}\\
  &\quad \quad \quad\quad \;\;{p_k} \geqslant 0,\forall k \in \mathcal{K}, \label{0-3} \\
  &\quad \quad\quad \quad \;\;\varphi _m^n = \frac{{i_m^n}}{{{2^{b}}}}\pi ,\forall n \in \mathcal{N},\forall m \in \mathcal{M}, \label{0-4} \\
  &\quad \quad\quad \quad \;\;{\log _2}\left( {1 + {\gamma _k}} \right) \geqslant \gamma _k^{\min },\forall k \in \mathcal{K}, \label{0-5} 
  \end{align}
\end{subequations}
where ${\eta _k}\left( {0 \leqslant {\eta _k} \leqslant 1} \right)$ represents the weight of the $k$-th Bob. $P_A$ in constraint \eqref{0-2} denote the max transmit power budget of Alice and the constraint \eqref{0-3} is the power variable constraint. Then, constraint \eqref{0-4} accounts for the limitation on the phase shifter with $b$-bit resolutions for individual transmission meta-atoms of the SIM. Also, constraint \eqref{0-5} ensures that the achievable rate of legitimate users cannot be lower than the preset minimum required user rate $\gamma_{k}^{\rm{min}}$. 

Note that the design objective function \eqref{0-1} is generally non-convex, and the discrete phase shift constraint \eqref{0-4} of SIM is also non-convex. Furthermore, the SIM phase shifts and Alice power coefficients are highly intertwined in the objective function, making it intractable to acquire the optimal solution for problem \eqref{P_0}.
On the other hand, due to the fact that SIM is composed of multiple metasurface layers, each layer consists of a large number of meta-atoms, such that the dimension of its solution space is generally tremendous. As a result, learning based methods are suitable for optimizing such complex, high-dimensional, and multivariate problems. In the next section, we will provide an efficient learning-based algorithm framework for solving the complex problem.

\section{Learning-Based Optimization Framework}\label{Learning_framework}
\begin{figure*}[!t]
\centering
\includegraphics[scale=0.29]{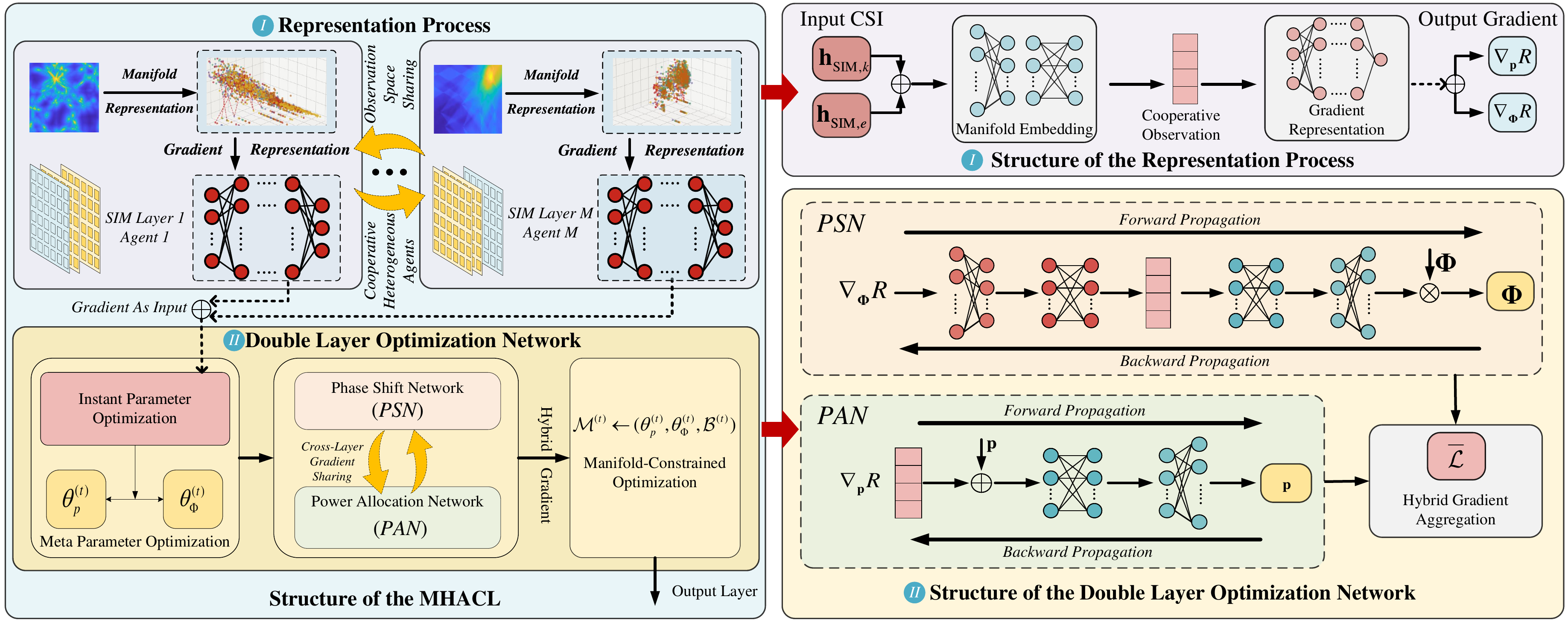}
\caption{Networking architecture of the proposed MHACL framework.}\label{networking} \vspace{-0.2cm}
\end{figure*}

This section presents the \textit{manifold heterogeneous agent continual learning} (MHACL), a model-agnostic optimization framework designed to address and mitigate suboptimal convergence in stochastic policy learning for cooperative multi-agent systems \cite{zhu2024marl}. The framework, as shown in Fig. \ref{networking}, integrates three principal components: 1) a manifold-based optimization theory for SIM-assisted multi-agent coordination derived in Section \ref{sec:manifold}, 2) a continual learning mechanism with dynamic policy adaptation detailed in Section \ref{sec:hacl}, and 3) a unified low-complexity SIMHACL algorithmic template that synthesizes theoretical guarantees with computationally efficient implementations, as formalized in Section \ref{sec:simhacl}. The synergistic integration of these components establishes a mathematically grounded paradigm for addressing heterogeneous agent coordination while maintaining temporal consistency across evolving environments.

\subsection{Manifold Theory for SIM-Assisted Multi-Agent Systems}\label{sec:manifold}  
Conventional optimization approaches applied to SIM-assisted multi-agent systems typically face critical computational bottlenecks. Existing methods, including alternating optimization and gradient-based techniques, necessitate high-dimensional matrix operations (e.g., eigenvalue decomposition, gradient projection) to update the phase shift matrices $\{\mathbf{\Phi}_m\}$ at each iteration \cite{zhang2025marl}. This induces a superlinear computational complexity scaling $\mathcal{O}(MN^{1+\epsilon})$ with respect to metasurface layers $M$ and meta-atoms per layer $N$, rendering large-scale SIM implementations computationally prohibitive. While learning-based alternatives can employ neural networks to implicitly model phase shift responses, the combination of discrete phase constraints and multi-layer SIM architecture leads to an exponential parameter space dimensionality $\mathcal{O}(2^{MN})$. Consequently, these methods suffer from two critical limitations: 1) the training processes demand exponentially growing batch sizes $\mathcal{O}(2^{MN})$ to mitigate potential overfitting, and 2) gradient-based optimization becomes unstable due to vanishing Lipschitz constants in high-dimensional spaces \cite{zhu2024joint,liu2025onboard}.  

To overcome these fundamental limitations, we reformulate the optimization problem \eqref{P_0} by leveraging differential manifold theory. Before proceeding further, a pivotal property emerging from our analysis is provided:  

\textbf{Proposition 1 (Power Constraint Saturation).} For any nontrivial stationary point $\{p_k^*\}$ of Problem \eqref{P_0} with and achievable secure rate $R_c > 0$, the total power constraint is always active $\sum_{k=1}^K p_k^* = P_A$.  

\textit{Proof}: Please refer to Appendix \ref{appendix1}.  

This proposition eliminates the necessity for conventional bisection-based power allocation methods by establishing that the optimal solution at maximum transmit power. Building on this foundation, we derive the manifold structure governing SIM configurations:  

\textbf{Proposition 2 (Product Manifold Embedding).} The phase shift optimization over $M$ SIM layers resides on a structed product manifold $\mathbb{M} \triangleq \prod_{m=1}^M \mathbb{T}^{N_m}$, where $\mathbb{T}^{N_m} \cong [0, 2\pi)^{N_m}$ denotes the $N_m$-torus for the $m$-th layer of the SIM and $N_m$ is the dimenson of the torus. This manifold embedding implicitly enforces unit-modulus constraint $|\phi_m^n| = 1$, while reducing the search space dimensionality from $2\sum_{m=1}^M N_m$ to $\sum_{m=1}^M N_m$.  

\textit{Proof}: Please refer to Appendix \ref{appendix2}.  

The established product manifold structure guarantees two key advantages: 1) Riemannian gradient descent on $\mathbb{M}$ ensures convergence to a local maxima under the Lipschitz continuity of the secure rate function, and 2) the intrinsic dimension reduction mitigates the curse of dimensionality inherent in conventional approaches. The proposed MHACL framework exploits this geometric insight by parameterizing the phase-shift matrices through a low-dimensional manifold projection network (will be detailed in Section \ref{PSN}), bypassing direct optimization in the original $\mathcal{O}(MN)$-dimensional space. A comprehensive theoretical analysis will be provided in Section \ref{sec:simhacl} demonstrating that this approach achieves $\mathcal{O}(\sqrt{MN})$ complexity reduction while maintaining $\epsilon$-optimality in secure rate maximization.  

\subsection{Heterogeneous Agent Continual Learning}\label{sec:hacl}
Deploying manifold theory in practical multi-agent SIM systems requires handling high-dimensional optimization, non-stationary wireless environments, and SIM realizability constraints. To address these challenges, we introduce a model-free framework that unifies geometric deep learning with adaptive continual learning mechanisms. A decentralized actor–critic architecture operates on the product manifold $\mathbb{M} \triangleq \mathbb{M}_{\mathbf{p}} \times \mathbb{T}^{N_m}$ (Section \ref{sec:manifold}), ensuring that the learned policy $\pi(\mathbf{s})$ satisfies SIM physical constraints. The global objective is to maximize a Riemannian-constrained secure rate:
\begin{align}
\max_{\pi \in \Pi} \;\;& \mathbb{E}\Big[ R_c(\mathbf{p},\mathbf{\Phi}) \Big] \\
\text{s.t.}\;\;& \mathbf{p} \in \mathbb{M}_{\mathbf{p}}, \quad \mathbf{\Phi}\in\mathbb{T}^{N_m}.
\end{align}
Note that dynamic regularization and prioritized experience buffers maintain knowledge across environment shifts, while a dual-scale meta-network coordinates agent decisions and SIM reconfiguration through Riemannian gradient flows:
\begin{align}
\mathrm{grad}_{\mathbb{M}} \mathcal{J}(\pi) = \mathcal{P}_{\mathbb{M}} \big( \nabla \mathcal{J}(\pi) \big),
\end{align}
with $\mathcal{P}_{\mathbb{M}}(\cdot)$ denoting the manifold projection operator. This hierarchical design guarantees policy improvement while preserving EM constraints.

\subsubsection{Cooperative Heterogeneous Agent System}
Within the proposed architecture, each SIM layer acts as an autonomous agent under a centralized training with decentralized execution (CTDE) paradigm. Traditional learning-based methods typically map observed CSI measurements, $\mathbf{h}_{\mathrm{SIM,k}}$ and $\mathbf{h}_{\mathrm{SIM,e}}$, directly to optimized power allocation $\mathbf{p}^*$ and phase-shift parameters $\mathbf{\Phi}^*$ through opaque policy networks:
\begin{align}
(\mathbf{p}^*, \mathbf{\Phi}^*) \approx f_\theta(\mathbf{h}_{\mathrm{SIM,k}}, \mathbf{h}_{\mathrm{SIM,e}}).
\end{align}
While effective in limited cases, such black-box mappings obscure the underlying parameter dynamics and ignore higher-order gradient relationships essential for EM–AI coordination.

To overcome these limitations, we reconceptualize the state–action mapping via differential optimization. Instead of raw CSI, the shared state is represented by gradient tensors derived from the secure rate objective:
\begin{align}
\mathbf{g}_p &\triangleq \nabla_{\mathbf{p}} R_c(\mathbf{p},\mathbf{\Phi}), \qquad
\mathbf{g}_\Phi \triangleq \nabla_{\mathbf{\Phi}} R_c(\mathbf{p},\mathbf{\Phi}).
\end{align}
Each agent generates manifold-constrained updates
\begin{align}
\Delta\mathbf{p} &= \mathcal{U}_{\mathbf{p}}\big(\mathbf{g}_p\big), \quad
\Delta\mathbf{\Phi} = \mathcal{U}_{\Phi}\big(\mathbf{g}_\Phi\big),
\end{align}
where $\mathcal{U}_{\mathbf{p}}$ and $\mathcal{U}_{\Phi}$ denote Riemannian update operators. As such, the agent policy is trained by minimizing a gradient-driven loss:
\begin{align}
\mathcal{J}(\pi) = 
\mathbb{E}\left[ \sum_{t=1}^{T} \mathcal{L}^{(t)}\big(\mathbf{g}_p^{(t)},\mathbf{g}_\Phi^{(t)}\big) \right],
\end{align}
which enhances interpretability by exposing optimization trajectories through gradient vector fields and obfuscates raw CSI for improved security \cite{zhu2025joint}.

\subsubsection{Continual Learning Architecture}\label{cla}
The wireless environment evolves over time, inducing distribution shifts that render classical sequential fine-tuning ineffective. We therefore embed continual adaptation into a nested optimization structure that maintains historical knowledge while adapting to new channel conditions.

Let $\theta_p$ and $\theta_\Phi$ denote parameters of the power allocation and phase shift networks (PAN/PSN), respectively. For each environmental state $t$, localized adaptation follows
\begin{align}
\mathbf{p}^{(t)} &= \mathrm{PAN}\!\left(\mathbf{p}^{(0)}, \mathbf{\Phi}^{(t-1)}; \mathbf{g}_p^{(t)}\right), \\
\mathbf{\Phi}^{(t)} &= \mathrm{PSN}\!\left(\mathbf{p}^{(t)}, \mathbf{\Phi}^{(0)}; \mathbf{g}_\Phi^{(t)}\right),
\end{align}
where $\mathbf{\Phi}^{(t-1)}$ transfers prior EM knowledge and $\mathbf{p}^{(0)}$ is freshly initialized to avoid overfitting.

Knowledge consolidation is enforced by a hierarchical loss balancing immediate performance and long-term stability:
\begin{align}
\mathcal{L}^{(t)} &= \mathbb{E}[R_c^{(t)}] 
+ \lambda \sum_{k=1}^{t-1} \omega_k \mathcal{D}\big(\theta^{(t)},\theta^{(k)}\big),
\end{align}
where $\mathcal{D}(\cdot)$ measures the Riemannian distance between current and historical parameters, e.g.,
\begin{align}
\mathcal{D}(\theta^{(t)},\theta^{(k)}) 
= \big\|\log\big((\theta^{(k)})^{-1/2} \theta^{(t)} (\theta^{(k)})^{-1/2}\big)\big\|_F^2.
\end{align}

The meta-optimization layer updates network parameters via
\begin{align}
\theta_p^{(t+1)} &= \theta_p^{(t)} - \alpha_p \nabla_{\theta_p}\Big[ \mathcal{L}^{(t)} + \beta \mathcal{L}_{\mathrm{traj}}^{(t)} \Big], \\
\mathcal{L}_{\mathrm{traj}}^{(t)} &= -\log p\!\left(\Delta\mathbf{\Phi}^{(t)} \mid \{\Delta\mathbf{\Phi}^{(k)}\}_{k< t}\right),
\end{align}
where $\mathcal{L}_{\mathrm{traj}}$ preserves the statistical consistency of optimization trajectories.

\subsubsection{Power Allocation Network (PAN)}
The PAN implements manifold-constrained optimization through dynamic gradient assimilation. At each inner iteration $(i,j)$, the WSSR gradient
\begin{align}
\mathbf{g}_p^{(i,j)} = \nabla_{\mathbf{p}} R_{\mathbf{p}}^{(i,j)}
\end{align}
is refined via a masked neural transformation:
\begin{align}
\widetilde{\mathbf{p}}^{(i,j)} &= \mathbf{p}^{(i,j)} + \mathcal{N}_{\theta}\!\left(\Delta\mathbf{p}^{(i,j)} \odot \mathbf{M}^{(t-1)}\right), \label{eq:masked_update}
\end{align}
where $\mathbf{M}^{(t-1)}$ is a soft parameter mask preserving prior optimization patterns. The normalized update enforces instantaneous power constraints:
\begin{align}
\mathbf{p}^{(i+1,j)} =
\sqrt{\frac{P_A}{\mathrm{Tr}\!\big(\widetilde{\mathbf{p}}^{(i,j)} (\widetilde{\mathbf{p}}^{(i,j)})^H\big)}}\;
\widetilde{\mathbf{p}}^{(i,j)}. \label{eq:proj_norm}
\end{align}
Parameter updates follow a Riemannian Adam scheme:
\begin{align}
\theta_p^{(t)} = \theta_p^{(t-1)} - \alpha_p \,\mathrm{Adam}\big(\mathrm{grad}_{\mathbb{M}_{\mathbf{p}}}\overline{\mathcal{L}}\big).
\end{align}

\begin{algorithm}[t]
\caption{MHACL Process}
\label{alg:mhacl}
\begin{algorithmic}[1]
\Procedure{MHACL}{$\mathbf{h}_{\mathrm{e}}^{(t)}, \mathbb{M}^{(t-1)}$}
    \State Initialize $\theta_p^{(t)}, \theta_\Phi^{(t)} \gets \mathbb{M}^{(t-1)}$
    \State $\mathbf{p}^{(0)} \gets \sqrt{P/K}\mathbf{1}_K$, $\mathbf{\Phi}^{(0)} \gets \mathcal{P}_{\mathbb{T}^{MN}}(\mathbf{\Phi}^{(t-1)})$
    \State Initialize experience buffer $\mathcal{B}^{(t)} \gets \{\nabla R_c^{(t-1)}\}$
    \For{epoch $k \in 1:N_e$}
        \State Initialize loss $\overline{\mathcal{L}} \gets 0$
        \For{outer iteration $j \in 1:N_o$}
            \State Sample task batch $\{\nabla R_c^{(k)}\}_{k=1}^t \sim \mathcal{B}^{(t)}$
            \State $\mathbf{\Phi}^{(j,0)} \gets \mathbf{\Phi}^{(j-1,N_i)}$
            \For{inner iteration $i \in 1:N_i$}
                \State $\Delta\mathbf{\Phi} \gets \mathrm{PSN}(\nabla_{\Phi}R_c^{(t)} \odot \mathbf{M}_\Phi^{(t-1)})$
                \State Update $\mathbf{\Phi}^{(j,i)}$
                \State $\Delta\mathbf{p} \gets \mathrm{PAN}(\nabla_p R_c^{(t)} \odot \mathbf{M}_p^{(t-1)})$
                \State Update $\mathbf{p}^{(j,i)}$
            \EndFor
            \State $\mathcal{L}_j \gets -R_c(\mathbf{p}^*, \mathbf{\Phi}^*) + \lambda\mathcal{R}(\theta^{(t)},\theta^{(t-1)})$
            \State $\overline{\mathcal{L}} \gets \overline{\mathcal{L}} + \mathcal{L}_j$
            \State \textbf{Update buffer}: $\mathcal{B}^{(t)} \gets \mathcal{B}^{(t)} \cup \{\nabla R_c^{(j)}\}$
        \EndFor
        \State $\theta_p^{(t)} \gets \theta_p^{(t-1)} - \alpha_p\mathrm{Adam}(\nabla_{\theta_p}\overline{\mathcal{L}})$
        \If{$k \mod N = 0$}
            \State $\theta_\Phi^{(t)} \gets \theta_\Phi^{(t-1)} - \alpha_\Phi\mathrm{Adam}(\nabla_{\theta_\Phi}\overline{\mathcal{L}})$
        \EndIf
    \EndFor
    \State Update manifold memory $\mathbb{M}^{(t)} \gets (\theta_p^{(t)}, \theta_\Phi^{(t)}, \mathcal{B}^{(t)})$
    \State \Return $\mathbf{p}_{opt}, \mathbf{\Phi}_{opt}, \mathbb{M}^{(t)}$
\EndProcedure
\end{algorithmic}
\end{algorithm}

\subsubsection{Phase Shift Network (PSN)}\label{PSN}
The PSN must respect the toroidal geometry of phase parameters. A frequency-aware update mechanism refines the phase gradient
\begin{align}
\Delta\widetilde{\phi} &= \lambda\,\sigma\!\left(\mathcal{T}_{\phi}\big(\mathbf{g}_\Phi^{(i,j)} \odot \mathbf{C}^{(t-1)}\big)\right),
\end{align}
where $\mathcal{T}_{\phi}$ is a spectral transformer and $\mathbf{C}^{(t-1)}$ captures historical interference patterns. Phase updates are expressed as rotations on the unit circle:
\begin{align}
\mathbf{\Phi}^{(i+1,j)} &= \mathrm{diag}\!\left[ e^{j(\mathbf{\Phi}^{(i,j)} + \Delta\widetilde{\phi})} \right],
\end{align}
ensuring continuous evolution on the manifold $\mathbb{T}^{N_m}$. Parameter updates employ the same Riemannian Adam rule as PAN:
\begin{align}
\theta_\Phi^{(t)} = \theta_\Phi^{(t-1)} - \alpha_\Phi \,\mathrm{Adam}\big(\mathrm{grad}_{\mathbb{T}^{N_m}}\overline{\mathcal{L}}\big).
\end{align}

The spectral transformer enables cross-task transfer by encoding recurrent interference patterns into orthogonal basis functions, preventing catastrophic forgetting and promoting rapid adaptation to new SIM setups.

\subsection{SIMHACL: Low-Complexity Continual Learning Template}\label{sec:simhacl}
To further reduce computational burden while preserving adaptability, we develop a simplified continual learning scheme named \textbf{SIMHACL}, which embeds EM constraints into a product-manifold representation, compressing the high-dimensional phase-shift search into a compact parameter subspace. The unit-modulus condition is automatically satisfied through Riemannian gradient flows, thereby eliminating explicit projection steps and guaranteeing physical feasibility.

For transmit power control, SIMHACL exploits the nearly orthogonal structure of optimal allocations under full-power saturation (Proposition 1). Starting from an equal-power initialization, gradient-driven refinements adaptively approach the saturation boundary:
\begin{align}
\mathbf{p}^{(t+1)} = \mathcal{P}_{\mathbb{M}_{\mathbf{p}}}\big[\mathbf{p}^{(t)} + \eta \nabla_{\mathbf{p}} R_c(\mathbf{p}^{(t)},\mathbf{\Phi}^{(t)}) \big],
\end{align}
where $\mathcal{P}_{\mathbb{M}_{\mathbf{p}}}$ denotes the Riemannian projection on the power manifold and $\eta$ is a step-size parameter.

The overall algorithm achieves \emph{linear} complexity scaling $\mathcal{O}\big(M \max N_m\big)$ by combining two key mechanisms. First, gradient preconditioning employs learned preconditioners to replace costly matrix inversions, which preserves numerical stability 
and removes the dominant cubic operations present in conventional solvers. Then, manifold embedding compresses parameter updates into intrinsic manifold coordinates, effectively reducing redundant dimensions and ensuring that each update remains within the feasible geometric space.

These structural simplifications lead to a clear hierarchy of computational costs across different design frameworks. Conventional alternating optimization suffers from cubic growth with respect to the number of metasurface meta-atoms, requiring a complexity of 
$\mathcal{O}\big(K(L M^3 + L^2 N^2)\big)$. By leveraging decentralized gradient flows, the proposed MHACL decouples the cubic term and reduces the cost to $\mathcal{O}(L M N)$, which already enables efficient deployment in moderately large systems. Building on this foundation, SIMHACL further advances the reduction by achieving linear complexity in both $M$ and $N$, lowering the quadratic dependence in each iteration to a purely linear scaling and 
rendering the algorithm effectively insensitive to the number of antennas or meta-atoms.

\section{Numerical Results}\label{simulation}

\subsection{Simulation Setup}
This section presents numerical results to analyze the proposed algorithms and evaluate the performance of the considered SIM-based multi-user MIMO system. The simulation settings are given as follows. A BS serves $K$ single-antenna users and we assume that the distance between the BS and the UE cluster center is $30\,\text{m}$, while each UE experiences a random movement within a range of $10\,\text{m}$ with respect to the center. The eavesdropper is located at the center point of the user cluster. The height of the SIM-assisted BS, users, and eavesdropper is 15 $\rm{m}$ and 1.65 $\rm{m}$, 1.65 $\rm{m}$, respectively. 
Furthermore, we assume that the thickness of the SIM is $T_{\rm{SIM}} = 5\lambda$, ensuring that the spacing between two adjacent metasurfaces for an $M$-layer SIM is ${d_{{\rm{Layer}}}} = {{{T_{{\rm{SIM}}}}} \mathord{\left/
 {\vphantom {{{T_{{\rm{SIM}}}}} M}} \right.
 \kern-\nulldelimiterspace} M}$. 
Furthermore, we focus on a square metasurface arrangement with $N={N_x}{N_y}$ meta-atoms where $N_x = N_y$, and $N_x$ and $N_y$ denote the number of meta-atoms along the $x$-axis and $y$-axis, respectively. Furthermore, we assume half-wavelength spacing between adjacent antennas/meta-atoms at the BS and metasurface layers. Also, the size of each meta-atom is $d = {d_x} = {d_y} = {\lambda  \mathord{\left/
 {\vphantom {\lambda  2}} \right.
 \kern-\nulldelimiterspace} 2}$.
We assume a correlated Rayleigh fading channel model, and the distance-dependent path loss is modeled as 
 \begin{align}
      {\beta _{k}} = {C_0}{\left( {{{{d_{k}}} \mathord{\left/
 {\vphantom {{{d_{lk}}} {{d_0}}}} \right.
 \kern-\nulldelimiterspace} {{d_0}}}} \right)^{\varpi_{0} }},\quad {d_{k}} > {d_0},
 \end{align}
where $d_{k}$ denotes the link distance between the SIM-assisted BS to the $k$-th UE. Also, ${C_0} = {\left( {{\lambda  \mathord{\left/
 {\vphantom {\lambda  {4\pi {d_0}}}} \right.
 \kern-\nulldelimiterspace} {4\pi {d_0}}}} \right)^2}$ denotes the free space path loss with respect to the reference distance $d_0 = 1$ m \cite{rappaport2015wideband}, and ${\varpi_{0}} =-3.5$ denotes the path loss exponent. Besides, we consider a system operating at a carrier frequency of 28 GHz with a transmission bandwidth of 10 MHz, and an effective noise power spectral density of $-174$ dBm/Hz.
The simulation results are obtained by averaging over 100 independent experiments.

\subsection{Algorithm Convergence}

\begin{figure}[t]
\centering
\includegraphics[scale=0.375]{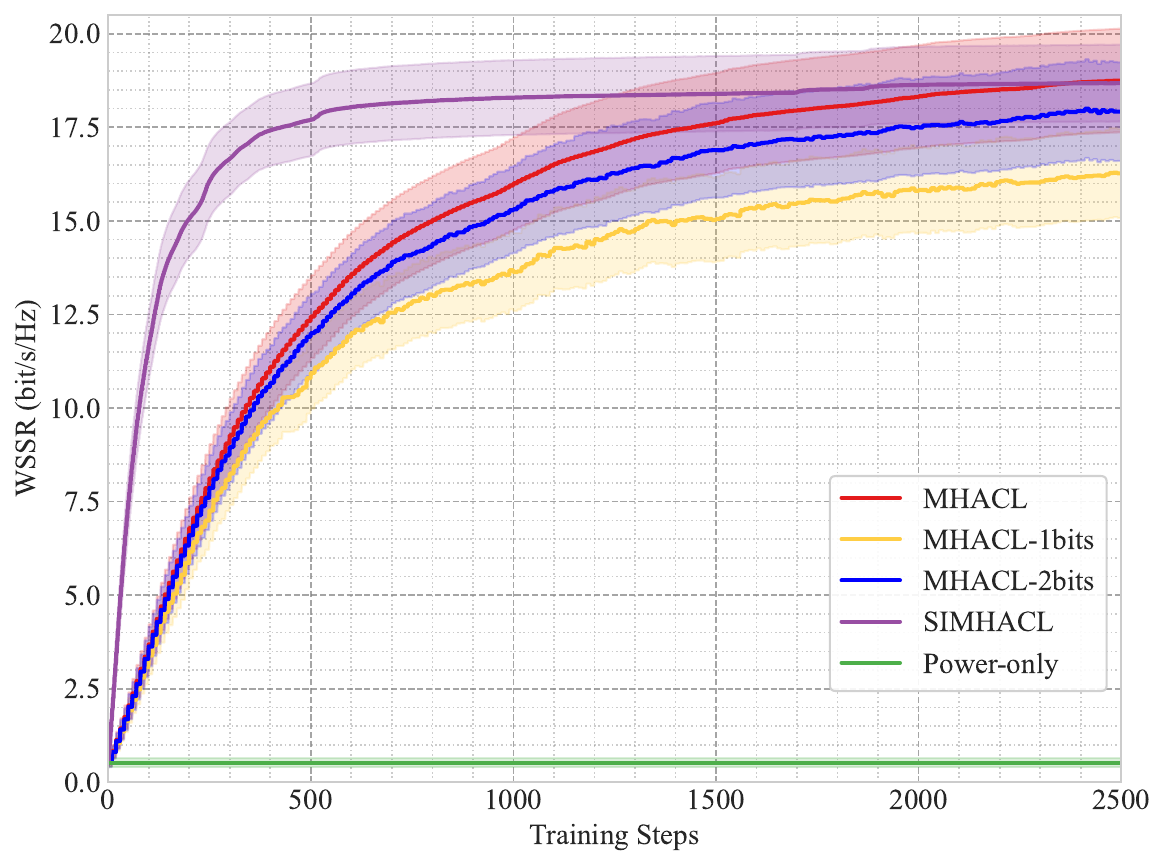}
\caption{Convergence rate with discrete phase over different schemes ($L = 4$, $M = 4$, $K = 4$, $N = 64$).} \label{Fig_conv}
\end{figure}
As shown in Fig.~\ref{Fig_conv}, the proposed MHACL algorithmic framework demonstrates rapid convergence under both continuous and discrete phase shift scenarios. In particular, the proposed low-complexity SIMHACL algorithm converges within 500 iterations, significantly faster than the MHACL algorithm, which requires 2000 iterations. Moreover, the performance degradation of SIMHACL after convergence is minimal compared to MHACL, is indicated by the 95\% confidence interval shading in Fig.~\ref{Fig_conv}. The reason is that the template achieves linear complexity scaling $\mathcal{O}(M\max N_m)$, rather than the cubic computational complexity of MHACL, by sidestepping matrix inversions with learned gradient preconditioners and further reducing parameter updates through manifold embeddings. Additionally, the orthogonal power allocation on SIM antennas ensures WSSR performance.

\begin{figure}[t]
\centering
\includegraphics[scale=0.375]{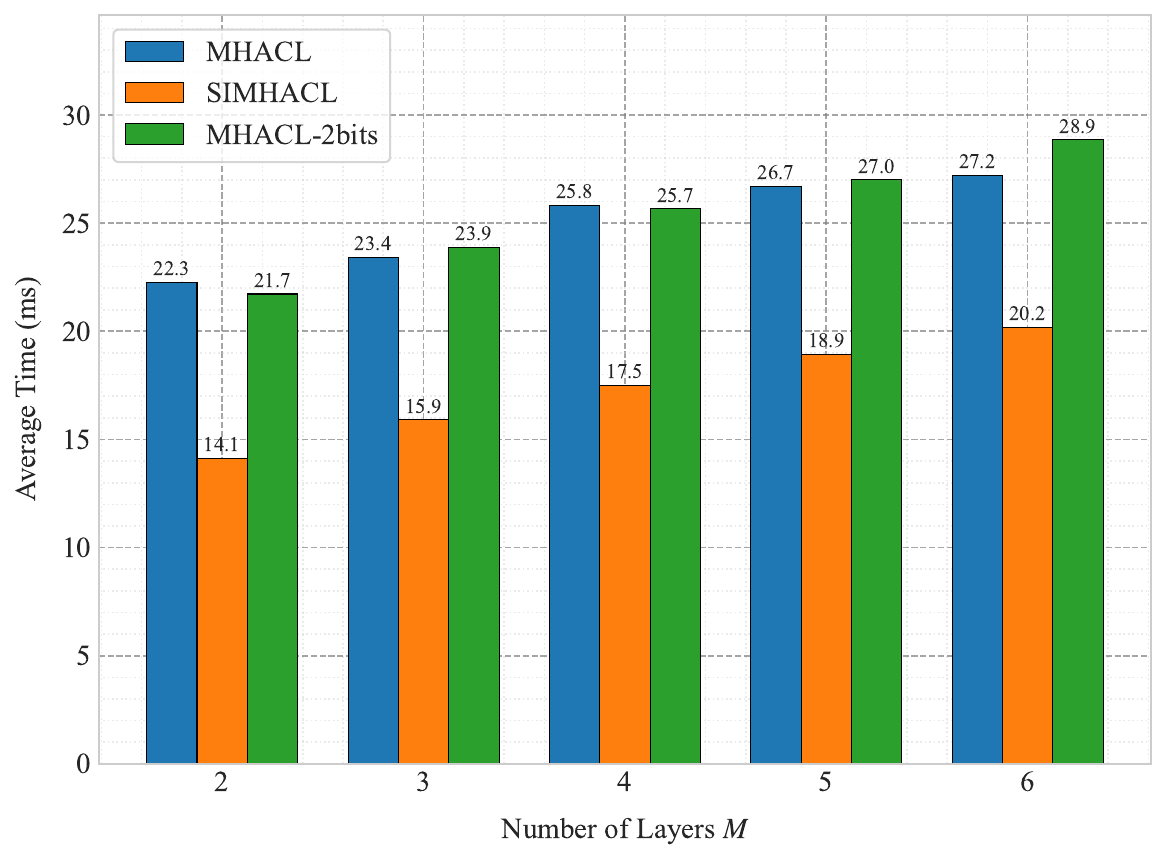}
\caption{Average computing time against the number of SIM layers with discrete phase over different schemes for one step ($L = 4$, $K = 4$, $N = 64$).} \label{Fig_time}
\end{figure}
Fig.~\ref{Fig_time} illustrates the average computation time against the number of SIM layers by considering various algorithms for one training step. It is clear that the proposed framework achieves millisecond-level training time per iteration for SIM-assisted systems. Notably, as the number of SIM layers increases, the required computational complexity exhibits a proportional rise; however, the training overhead remains manageable despite exponential growth in optimization parameters, demonstrating the adaptability of the proposed framework. Furthermore, the low-complexity SIMHACL algorithm demonstrates a 30\% reduction in training time compared to MHACL. Furthermore, discrete phase shift configurations maintain comparable latency profiles to their continuous counterparts, underscoring the framework’s generality and robustness to practical implementation constraints.

\subsection{Impact of System Parameters}

\begin{figure}[t]
\centering
\includegraphics[scale=0.375]{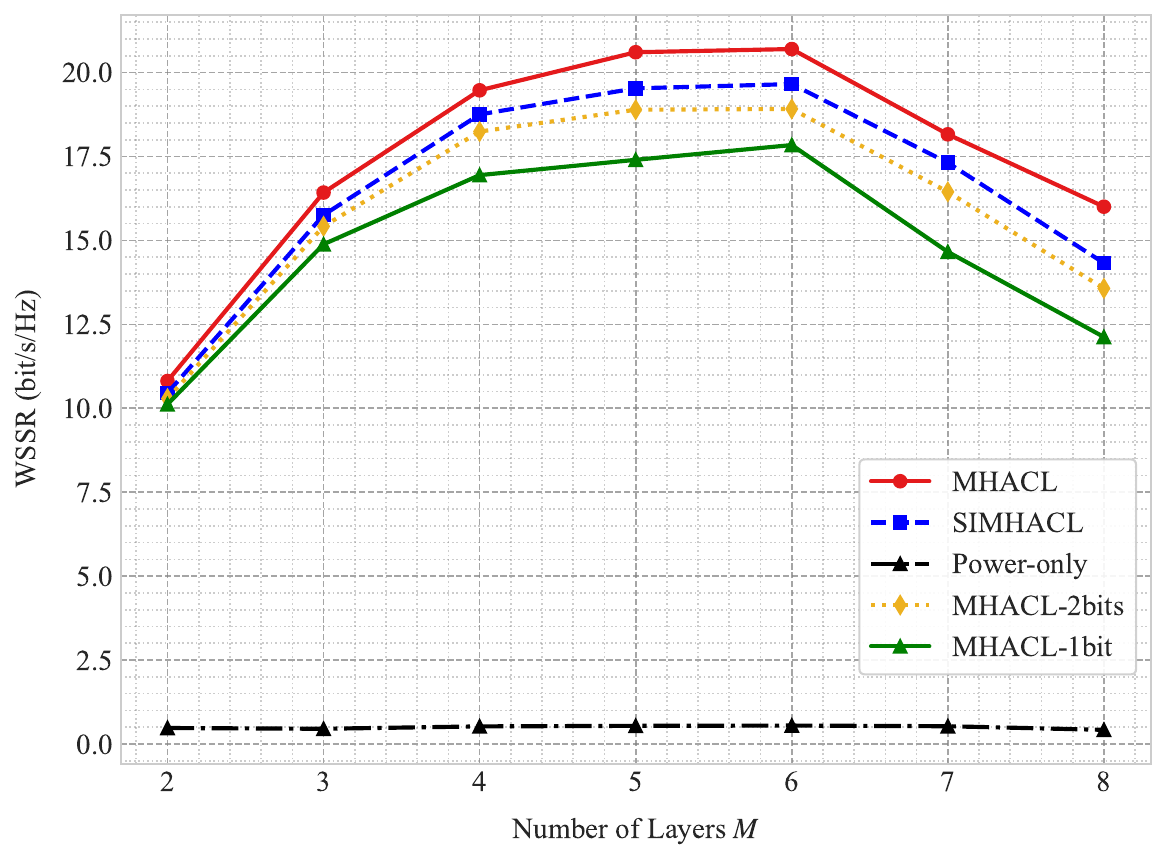}
\caption{The WSSR against the number of SIM layers with discrete phase over different schemes ($L = 4$, $K = 4$, $N = 64$).} \label{Fig_num_layer}
\end{figure}
Fig.~\ref{Fig_num_layer} illustrates the WSSR performance of different algorithms as a function of the number of SIM layers. The MHACL algorithm achieves the best performance among all considered schemes. The results show that when the number of SIM layers is small, increasing the number of SIM layers significantly improves system performance. For example, the WSSR performance improves by roughly 86\% with a 6-layer SIM compared to a 2-layer configuration. However, when the number of layers exceeds a certain threshold ($M = 6$), further increasing it does not continue to enhance the WSSR and may even degrade performance. This is because when the number of SIM layers is small, adding more layers increases the accuracy and degrees of freedom in signal processing by introducing additional meta-atoms. When the number of layers becomes large, the inter-layer transmission fading becomes dominant as shown in \eqref{G}, thus resulting in performance degradation. 
In contrast, the Power-only scheme performs significantly worse across all cases, highlighting the importance of SIM wave-based beamforming in enhancing physical layer security.

\begin{figure}[t]
\centering
\includegraphics[scale=0.375]{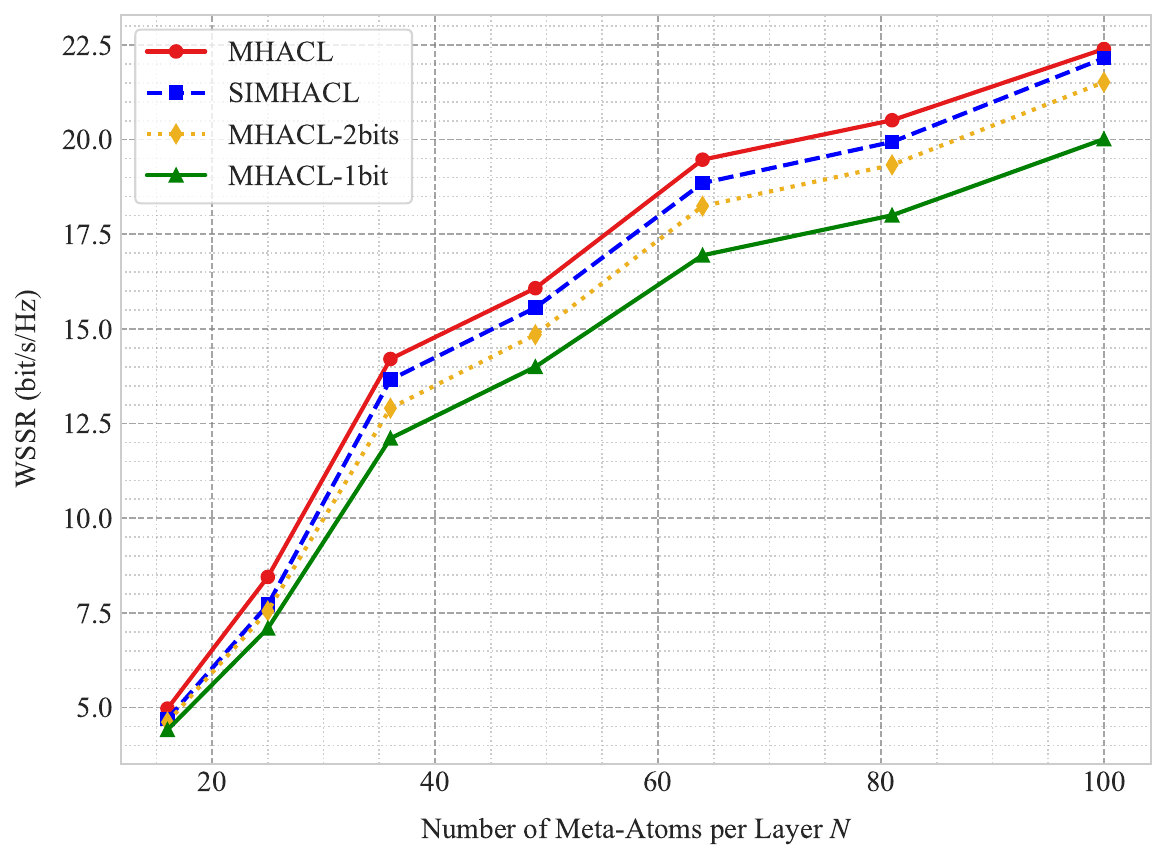}
\caption{The WSSR against the number of meta-atoms per layer with discrete phase over different schemes ($L = 4$, $M = 4$, $K = 4$).} \label{Fig_num_atom}
\end{figure}
Fig.~\ref{Fig_num_atom} illustrates the WSSR performance of different algorithms as a function of the number of SIM meta-atoms per layer. It can be observed that, due to the increased spatial degrees of freedom, the WSSR improves with the increase of $N$ across all schemes. For instance, compared to the case with 16 meta-atoms per layer, a SIM with 64 meta-atoms per layer achieves a threefold improvement in WSSR performance, thanks to the larger metasurface aperture.
Moreover, the proposed SIMHACL algorithm achieves performance close to that of the MHACL algorithm, demonstrating its efficiency. Although the MHACL-2bits and MHACL-1bit schemes exhibit performance degradation due to phase quantization, they still benefit from the increase in $N$.
In summary, increasing the number of SIM meta-atoms per layer is consistently a practical approach to enhancing the system secrecy rate. It is also worth noting that SIMHACL consistently outperforms both quantized-phase schemes, indicating the robustness and generality of the proposed framework under both continuous and discrete phase constraints.

\begin{figure}[t]
\centering
\includegraphics[scale=0.375]{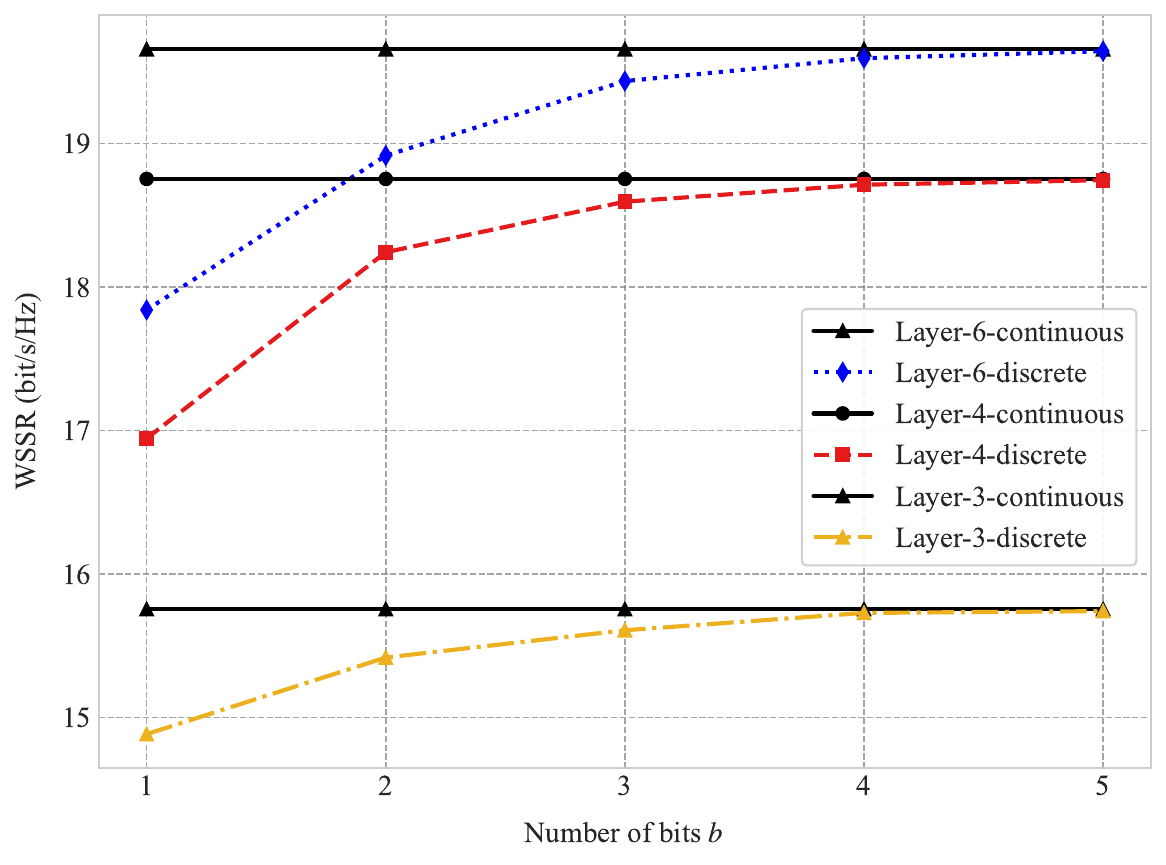}
\caption{The WSSR against the number of quantization bits of the phase shifts over different numbers of the SIM layers ($L = 4$, $K = 4$, $N = 64$).} \label{Fig_bit}
\end{figure}
Fig.~\ref{Fig_bit} illustrates the WSSR performance versus the number of quantization bits $b$ under different SIM layer configurations, considering both continuous and discrete phase shift models.
It can be observed that the WSSR increases monotonically with higher phase resolution across all configurations. In particular, the performance gain is more significant in the low-bit regime (e.g., $b = 1 \sim 3$), highlighting the sensitivity of secrecy performance to phase quantization in low-resolution settings. 
In addition, SIM configurations with a greater number of layers consistently achieve higher WSSR across all quantization levels, demonstrating the effectiveness of increased layer depth in enhancing spatial diversity and secrecy capacity. However, when the phase is quantized with only 1 bit, the performance gap between discrete-phase and continuous-phase schemes reaches 10\% under the six-layer configuration, compared to just 6\% in the two-layer case. This suggests that the sensitivity of WSSR performance to phase resolution increases with the number of SIM layers, thereby necessitating finer quantization to maintain performance at higher layer depths.
As expected, continuous-phase schemes always outperform their discrete-phase counterparts for a given number of layers. Nevertheless, the performance gap diminishes by increasing the quantization resolution. When the quantization level reaches 4 bits, the discrete-phase performance closely approaches that of the continuous-phase case, indicating that high-resolution discrete implementations can effectively emulate ideal continuous-phase designs in practice.

\begin{figure}[t]
\centering
\includegraphics[scale=0.375]{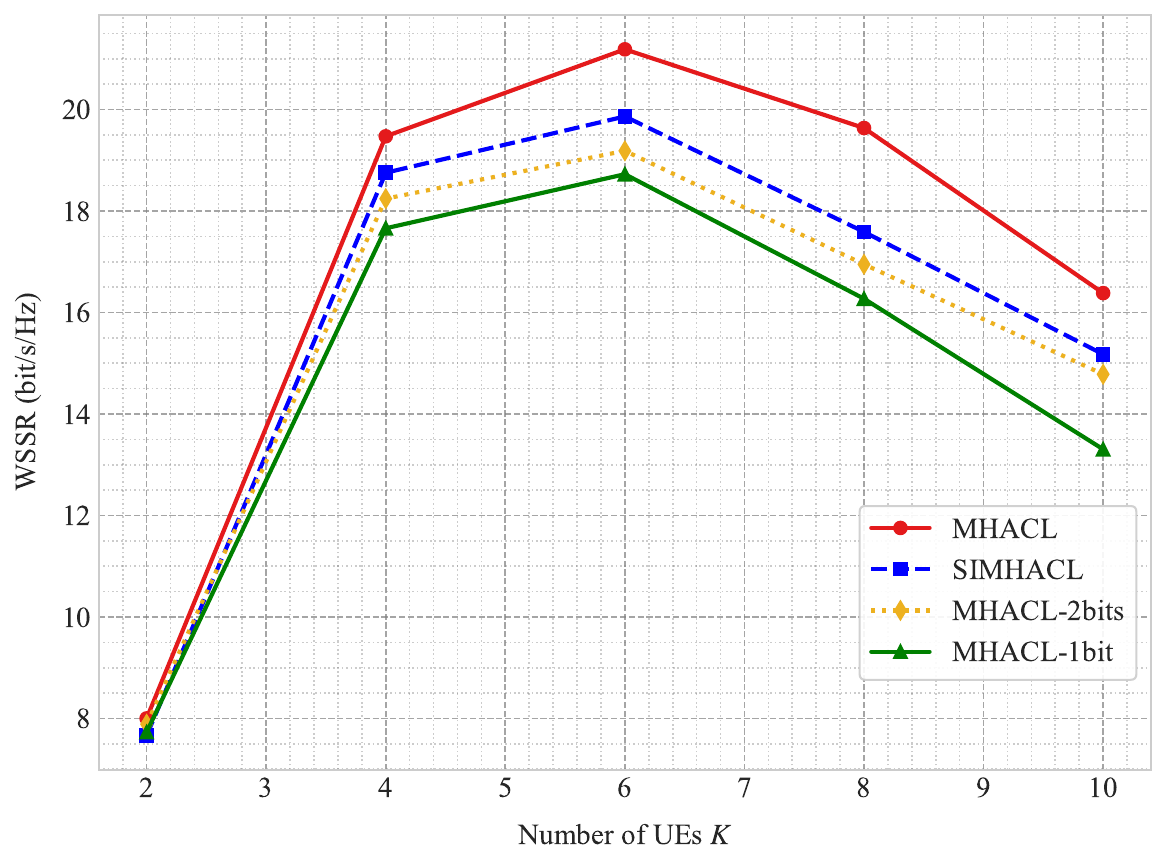}
\caption{The WSSR against the number of users with discrete phase over different schemes ($L = 4$, $M = 4$, $N = 64$).} \label{Fig_num_UE}
\end{figure}
Fig.~\ref{Fig_num_UE} illustrates the WSSR performance of different algorithms as a function of the number of users.
As observed, the WSSR performance across all schemes initially increases and then decreases with the growing number of users, which differs from the behavior in conventional MIMO systems. When the number of users is relatively small, the system performance does not reach saturation; thus, increasing the number of users can improve resource utilization and enhance overall system performance. However, as the number of users continues to grow, each antenna, restricted to transmitting a single-user data stream, receives less power under a fixed total power budget. Meanwhile, the increased multi-user interference further degrades the secrecy rate performance.
This indicates that in SIM-assisted systems, it is crucial to select the number of served users to avoid excessive multi-user interference, which leads to performance degradation.

\begin{figure}[t]
\centering
\includegraphics[scale=0.375]{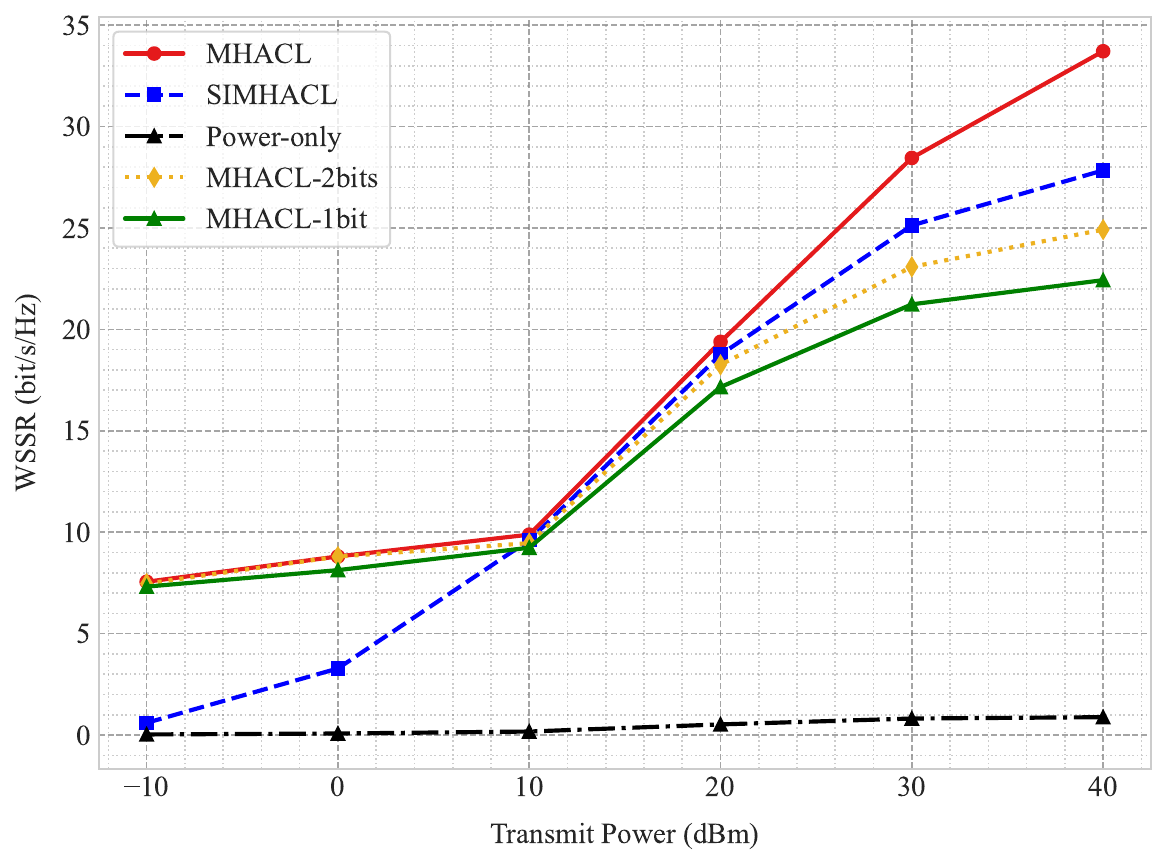}
\caption{The WSSR against different values of transmit power with discrete phase over different schemes ($L = 4$, $M = 4$, $K = 4$, $N = 64$).} \label{Fig_power}
\end{figure}
Fig.~\ref{Fig_num_UE} illustrates the WSSR performance of different algorithms as a function of the transmit power.
It is observed that for all schemes, the WSSR increases with the increase in transmit power. Moreover, as shown by the power-only scheme, when the SIM phases are randomly configured, increasing the transmit power results in only marginal performance improvement, highlighting the critical role of SIM wave-domain precoding.
Interestingly, under low transmit power conditions, the proposed SIMHACL scheme exhibits significantly worse performance compared to MHACL. However, as the transmit power increases, the performance of SIMHACL improves rapidly and approaches that of MHACL. This occurs because SIMHACL consistently adopts orthogonal power allocation. However, when the total antenna power is relatively low, orthogonality is not the optimal solution. In such situations, the antenna gain and board-to-board losses cancel each other out, resulting in SIMHACL performing significantly worse compared to MHACL. As the SIM antenna's total power rises, the optimal power allocation increasingly conforms to orthogonal eigenmodes, with the channel gain offsetting the insertion losses. Consequently, SIMHACL's performance significantly improves and approaches optimal conditions.

\section{Conclusion}
In this paper, we proposed a novel learning-based optimization framework for enhancing PLS in SIM-assisted multi-user MIMO secure communication systems. By introducing a SIM at the BS, we leveraged wave-based beamforming to enable per-antenna single-stream transmission with significantly reduced hardware complexity. A WSSR maximization problem was formulated under practical constraints, including discrete phase quantization and power limitations. To tackle the inherent non-convexity and high dimensionality, we developed a MHACL framework, which effectively handles joint power and phase shift optimization in dynamic environments. Furthermore, a low-complexity variant named SIMHACL was proposed, which embeds the phase optimization into a product manifold structure, achieving linear complexity scaling and maintaining performance robustness. In particular, the proposed framework achieves millisecond-level training time per iteration for SIM-assisted systems. Numerical simulations validated the superiority of the proposed approaches, showing that SIMHACL achieves near-optimal performance with 30\% lower computational overhead compared to MHACL. These results demonstrate the great potential of SIMs and learning-based methods for secure and efficient communication in next-generation wireless networks.

\begin{appendices}
\section{Proof of Proposition 1}\label{appendix1}
For any nontrivial stationary point $\mathbf{p}_k^{*}$, we can employ the Fritz-John (FJ) condition on the original system WSSR maximization problem \eqref{P_0} and derive: 
\begin{subequations}
\begin{align}
 \mu_0 \nabla_\mathbf{p} R_{c}(\mathbf{p}^*) &+ \mu \nabla_\mathbf{p}\left(\sum_{k=1}^K p_k - P_{A}\right) \!+\!\! \sum_{k=1}^K \nu_k \nabla_\mathbf{p}(-p_k) \!=\! 0, \label{FJ1}\\
& \mu_{0} \geq 0, \\
& \mu \geq 0, \\
& \nu_k \geq 0, \forall k \in \mathcal{K}, \\
& \mu_{0} + \mu + \sum \nu_{k} > \mathbf{0},
\end{align}
\end{subequations}
where $\mu_{0}$ is the multiplier of the WSSR function, $\mu \geq 0$ is the multiplier of the power constraint $\sum p_{k} \leq P_{A}$, and $\nu_k$ is the multiplier of non-negative power constraint $p_k \geq 0$.

For non-trivial solutions where $\exists k$ s.t. $p_k^* > 0$, the complementary slackness is $\nu_k p_k^* = 0$, thus we get $\nu_k = 0$ for all $k$. Hence, FJ conditions simplify to:
\begin{align}
    \mu_0 \nabla_\mathbf{p} R_{c}(\mathbf{p}^*) + \mu \nabla_\mathbf{p}(\sum_{k=1}^K p_k - P_{A}) = \mathbf{0}.
\end{align}
Now we would prove by contradiction that $\mu$ is strictly positive. First, we take the derivative of $R_{c,k}$ with respect to $p_{k}$ and yield \eqref{delta} at the top of the next page, where 
\begin{align}
\setcounter{equation}{23}
\alpha_k &\triangleq \left| \mathbf{h}_{\text{SIM},k}^{\text{H}} \mathbf{Gw}_1^k \right|^2,  \\
\beta_k &\triangleq \left| \mathbf{h}_{\text{SIM},e}^{\text{H}} \mathbf{Gw}_1^k \right|^2, \\
I_k &\triangleq \sum_{k' \neq k} p_{k'} \left| \mathbf{h}_{\text{SIM},k}^{\text{H}} \mathbf{Gw}_1^{k'} \right|^2 + \sigma_k^2,  \\
I_e &\triangleq \sum_{k' \neq k} p_{k'} \left| \mathbf{h}_{\text{SIM},e}^{\text{H}} \mathbf{Gw}_1^{k'} \right|^2 + \sigma_e^2. 
\end{align}

\newcounter{mytempeqncnt}
\begin{figure*}[t!]
\normalsize
\setcounter{mytempeqncnt}{1}
\setcounter{equation}{22}
\begin{align}\label{delta}
& \nabla_{\mathbf{p}_{k}}R_{c,k} 
    = \nabla_{\mathbf{p}_{k}} \left[\log_{2}\left(1+\gamma_{k}\right) -\log_{2}\left(1+\gamma_{k}^{e}\right) \right] \notag \\
    &= \nabla_{\mathbf{p}_{k}}\left[\log_{2}\left(1+\frac{p_k\left| \mathbf{h}_{{\text{SIM}},k}^{\text{H}}\mathbf{Gw}_1^k \right|^2}{\sum\nolimits_{k' \ne k}^K p_{k'}\left| \mathbf{h}_{{\text{SIM}},k}^{\text{H}}\mathbf{Gw}_1^{k'} \right|^2 + \sigma_k^2}\right) \right. 
     \left. -\log_{2}\left(1+\frac{p_k\left| \mathbf{h}_{{\text{SIM}},e}^{\text{H}}\mathbf{Gw}_1^k \right|^2}{\sum\nolimits_{k' \ne k}^K p_{k'}\left| \mathbf{h}_{{\text{SIM}},e}^{\text{H}}\mathbf{Gw}_1^{k'} \right|^2 + \sigma_e^2}\right)\right] \notag \\
    & = \nabla_{\mathbf{p}_{k}}\left[\log_{2}\left(\sum\nolimits_{k'=1}^K p_{k'}\left| \mathbf{h}_{{\text{SIM}},k}^{\text{H}}\mathbf{Gw}_1^{k'} \right|^2 + \sigma_k^2\right) \right. 
	 \left. - \log_{2}\left(\sum\nolimits_{k'=1}^K p_{k'}\left| \mathbf{h}_{{\text{SIM}},e}^{\text{H}}\mathbf{Gw}_1^{k'} \right|^2 + \sigma_e^2\right)\right] \notag \\
	& = \frac{1}{\ln 2}\left(\sum\nolimits_{k'=1}^K p_{k'}\left| \mathbf{h}_{{\text{SIM}},k}^{\text{H}}\mathbf{Gw}_1^{k'} \right|^2 + \sigma_k^2\right)^{-1} \cdot \left| \mathbf{h}_{{\text{SIM}},k}^{\text{H}}\mathbf{Gw}_1^{k} \right|^2 \notag \\
	 &- \frac{1}{\ln 2}\left(\sum\nolimits_{k'=1}^K p_{k'}\left| \mathbf{h}_{{\text{SIM}},e}^{\text{H}}\mathbf{Gw}_1^{k'} \right|^2 + \sigma_e^2\right)^{-1} \cdot \left| \mathbf{h}_{{\text{SIM}},e}^{\text{H}}\mathbf{Gw}_1^{k} \right|^2 \notag \\
	&= \frac{1}{\ln 2} \left( \frac{|\mathbf{h}_{\text{SIM},k}^\text{H} \mathbf{Gw}_1^k|^2}{\sum_{k'=1}^K p_{k'} |\mathbf{h}_{\text{SIM},k}^\text{H} \mathbf{Gw}_1^{k'}|^2 + \sigma_k^2} \right. 
	 - \left. \frac{|\mathbf{h}_{\text{SIM},e}^\text{H} \mathbf{Gw}_1^k|^2}{\sum_{k'=1}^K p_{k'} |\mathbf{h}_{\text{SIM},e}^\text{H} \mathbf{Gw}_1^{k'}|^2 + \sigma_e^2} \right) \notag \\
	&= \frac{1}{\ln 2} \cdot \frac{\alpha_k I_e - \beta_k I_k}{(p_k \alpha_k + I_k)(p_k \beta_k + I_e)},
\end{align}
\setcounter{equation}{15}
\hrulefill
\end{figure*}

Similarly, we have $\frac{\partial}{\partial p_k}\left(\sum_{i=1}^K p_i - P_{A}\right) = 1$. Thus, the FJ condition can be written as
\begin{subequations}
\begin{align}
	& \mu_{0}\cdot \frac{1}{\ln 2} \cdot \frac{\alpha_k I_e - \beta_k I_k}{(p_k \alpha_k + I_k)(p_k \beta_k + I_e)} + \mu \cdot 1 = 0, \\
	& \mu_{0}\cdot \frac{1}{\ln 2} \cdot \frac{\alpha_k I_e - \beta_k I_k}{(p_k \alpha_k + I_k)(p_k \beta_k + I_e)} = -\mu, \forall k \in \mathcal{K}.
\end{align}
\end{subequations}
The structural properties of the governing equation enforce a dual sign constraint: The left-hand component maintains non-negativity through the conditions $\mu_0 \geq 0$ and $\alpha_k I_e - \beta_k I_k \geq 0$, while the right-hand term $\mu \geq 0$ inherently assumes non-positive values. This inherent opposition necessitates an exhaustive analysis of two mutually exclusive scenarios.

\paragraph{Case 1: $\mu_0 = 0$} 
Requiring $\mu > 0$ to satisfy $\mu_0 + \mu > 0$ introduces fundamental contradictions. Substitution into the FJ conditions produces the degenerate result $\mu = 0$, thereby violating the positivity requirement. This inconsistency conclusively establishes the necessity of $\mu_0 > 0$ for physically meaningful solutions.

\paragraph{Case 2: $\mu_0 > 0$} 
Normalization by $\mu_0$ reveals the essential proportional relationship:
\begin{equation}
\frac{1}{\ln 2} \cdot \frac{\alpha_k I_e - \beta_k I_k}{(p_k \alpha_k + I_k)(p_k \beta_k + I_e)} = -\frac{\mu}{\mu_0}, \quad \forall k \in \mathcal{K}.
\end{equation}
The strictly positive left-hand side, guaranteed by non-trivial solution requirements, compels $\mu > 0$ to maintain equality. Through the complementary slackness condition $\mu\left(\sum p_k^* - P_{A}\right) = 0$, this directly yields the power allocation constraint:
\begin{equation}
\sum_{k=1}^K p_k^* = P_{A}. \quad \blacksquare
\end{equation}
Proposition 1 maintains validity throughout the learning optimization process despite frequent updates to $\mathbf{\Phi}$, as its proof relies on fundamental variational principles rather than transient parameter configurations.

\section{Proof of Proposition 2}\label{appendix2}
The manifold construction initiates by defining a composite parameter vector that consolidates phase shift parameters across all $M$ SIM layers, expressed as  
\begin{equation}  
\!\!\!\mathbf{\phi} \!=\! [\phi_{1,1},\ldots,\phi_{1,N_1},\ldots,\phi_{M,1},\ldots,\phi_{M,N_M}]^T \!\!\in\! \mathbb{R}^{\sum N_m}\!.  
\end{equation}  
This vector is embedded into the complex torus manifold through the mapping  
\begin{equation}  
\mathcal{E}(\mathbf{\phi}) = \left\{\mathrm{diag}\left(e^{j\phi_{m,1}},\ldots,e^{j\phi_{m,N_m}}\right)\right\}_{m=1}^M,  
\end{equation}  
where each diagonal matrix resides on $\mathbb{T}^{N_m}$, the $N_m$-dimensional complex torus. For the cascaded effective channel $\mathbf{H}_{\mathrm{eff}} = \prod_{m=1}^M \mathbf{\Phi}_m\mathbf{G}_m$, the Riemannian gradient computation involves layer-wise backward propagation. The gradient on manifold $\mathbb{M}_m$ takes the form  
\begin{equation}  
\mathrm{grad}_{\mathbb{M}_m} R_{c} = 2\mathrm{Im}\left(\mathbf{\Phi}_m^* \odot \frac{\partial R_{c}}{\partial \mathbf{\Phi}_m}\right),  
\end{equation}  
where the partial derivative $\partial R_{c}/\partial \mathbf{\Phi}_m$ is computed through the differentiation chain, which is denoted as  
\begin{align}  
\frac{\partial \mathbf{H}_{\mathrm{eff}}}{\partial \mathbf{\Phi}_m} &= \left(\prod_{k=m+1}^M \mathbf{\Phi}_k\mathbf{G}_k\right)\mathbf{G}_m  
\times \left(\prod_{k=1}^{m-1} \mathbf{\Phi}_k\mathbf{G}_k\right),  
\label{eq:cascade_deriv}  
\end{align}  
demonstrating how gradients propagate through upstream and downstream layers.  

The alternating optimization framework sequentially updates parameters using retraction operators that preserve manifold constraints. Each iteration applies  
\begin{align}  
\mathbf{\phi}_m^{(t+1)} &= \mathcal{P}_{\mathbb{M}_m}\left(\mathbf{\phi}_m^{(t)} + \alpha_m \mathrm{grad}_{\mathbb{M}_m} R\right),  
\label{eq:alt_update}  
\end{align}  
for all $m \in \{1,\ldots,M\}$, where $\mathcal{P}_{\mathbb{M}_m}$ ensures adherence to the complex torus geometry. Computational comparisons highlight the manifold approach's superiority: Conventional parameterization demands $\mathcal{O}(2MN)$ complex parameters with $\mathcal{O}(MN)$ non-convex constraints, while the proposed method reduces dimensionality by 50\% through real-valued encoding. Gradient backpropagation scales as $\mathcal{O}(M \cdot \max N_m)$ per iteration, avoiding projection operations that typically dominate constrained optimization.  

This geometric formulation guarantees local convergence under Lipschitz continuity of the reward function $R(\cdot)$, as established by Riemannian optimization theory. The complexity reduction factor remains invariant to the number of layers $M$, ensuring scalability for large SIM configurations. 

\end{appendices}

\bibliographystyle{IEEEtran}
\bibliography{IEEEabrv,Ref}

\begin{thebibliography}{10}
\providecommand{\url}[1]{#1}
\csname url@samestyle\endcsname
\providecommand{\newblock}{\relax}
\providecommand{\bibinfo}[2]{#2}
\providecommand{\BIBentrySTDinterwordspacing}{\spaceskip=0pt\relax}
\providecommand{\BIBentryALTinterwordstretchfactor}{4}
\providecommand{\BIBentryALTinterwordspacing}{\spaceskip=\fontdimen2\font plus
\BIBentryALTinterwordstretchfactor\fontdimen3\font minus \fontdimen4\font\relax}
\providecommand{\BIBforeignlanguage}[2]{{%
\expandafter\ifx\csname l@#1\endcsname\relax
\typeout{** WARNING: IEEEtran.bst: No hyphenation pattern has been}%
\typeout{** loaded for the language `#1'. Using the pattern for}%
\typeout{** the default language instead.}%
\else
\language=\csname l@#1\endcsname
\fi
#2}}
\providecommand{\BIBdecl}{\relax}
\BIBdecl

\bibitem{wang2023road}
C.-X. Wang, X.~You, X.~Gao, X.~Zhu, Z.~Li, C.~Zhang, H.~Wang, Y.~Huang, Y.~Chen, H.~Haas \emph{et~al.}, ``On the road to {6G}: Visions, requirements, key technologies and testbeds,'' \emph{IEEE Commun. Surveys Tuts.}, vol.~25, no.~2, pp. 905--974, Feb. 2023.

\bibitem{wang2024active}
D.~Wang, Z.~Wang, K.~Yu, Z.~Wei, H.~Zhao, N.~Al-Dhahir, M.~Guizani, and V.~C. Leung, ``Active aerial reconfigurable intelligent surface assisted secure communications: Integrating sensing and positioning,'' \emph{IEEE J. Sel. Areas Commun.}, vol.~42, no.~10, pp. 2769--2785, Oct. 2024.

\bibitem{sun2024secure}
R.~Sun, W.~Wang, L.~Xu, N.~Zhao, N.~Al-Dhahir, and X.~Wang, ``Secure beamforming for {IRS}-assisted {NOMA} {SWIPT} networks,'' \emph{IEEE Trans. Commun.}, vol.~72, no.~11, pp. 6796--6809, Nov. 2024.

\bibitem{xing2023joint}
J.~Xing, T.~Lv, W.~Li, W.~Ni, and A.~Jamalipour, ``Joint optimization of beamforming and noise injection for covert downlink transmissions in cell-free internet of things networks,'' \emph{IEEE Int. Things J.}, vol.~11, no.~6, pp. 10\,525--10\,536, Jun. 2023.

\bibitem{wang2024tutorial}
Z.~Wang, J.~Zhang, H.~Du, D.~Niyato, S.~Cui, B.~Ai, M.~Debbah, K.~B. Letaief, and H.~V. Poor, ``A tutorial on extremely large-scale {MIMO} for {6G}: Fundamentals, signal processing, and applications,'' \emph{IEEE Commun. Surveys Tuts.}, vol.~26, no.~3, pp. 1560--1605, Mar. 2024.

\bibitem{10556753}
E.~Shi, J.~Zhang, H.~Du, B.~Ai, C.~Yuen, D.~Niyato, K.~B. Letaief, and X.~Shen, ``{RIS}-aided cell-free massive {MIMO} systems for {6G}: Fundamentals, system design, and applications,'' \emph{Proc. IEEE}, vol. 112, no.~4, pp. 331--364, Apr. 2024.

\bibitem{wei2024star}
W.~Wei, X.~Pang, C.~Xing, N.~Zhao, and D.~Niyato, ``{STAR-RIS} aided secure {NOMA} integrated sensing and communication,'' \emph{IEEE Trans. Wireless Commun.}, vol.~23, no.~9, pp. 10\,712--10\,725, Sep. 2024.

\bibitem{wu2019intelligent}
Q.~Wu and R.~Zhang, ``Intelligent reflecting surface enhanced wireless network via joint active and passive beamforming,'' \emph{IEEE Trans. Wireless Commun.}, vol.~18, no.~11, pp. 5394--5409, Nov. 2019.

\bibitem{10679239}
Y.~Sun, Z.~Lin, K.~An, D.~Li, C.~Li, Y.~Zhu, D.~Wing Kwan~Ng, N.~Al-Dhahir, and J.~Wang, ``Multi-functional {RIS}-assisted semantic anti-jamming communication and computing in integrated aerial-ground networks,'' \emph{IEEE J. Sel. Areas Commun.}, vol.~42, no.~12, pp. 3597--3617, Dec. 2024.

\bibitem{li2025covert}
X.~Li, M.~Liu, S.~Dang, N.~C. Luong, C.~Yuen, A.~Nallanathan, and D.~Niyato, ``Covert communications with enhanced physical layer security in {RIS}-assisted cooperative networks,'' \emph{IEEE Trans. Wireless Commun.}, vol.~24, no.~7, pp. 5605--5619, Jul. 2025.

\bibitem{xiao2023simultaneously}
H.~Xiao, X.~Hu, P.~Mu, W.~Wang, T.-X. Zheng, K.-K. Wong, and K.~Yang, ``Simultaneously transmitting and reflecting {RIS} ({STAR-RIS}) assisted multi-antenna covert communication: Analysis and optimization,'' \emph{IEEE Trans. Wireless Commun.}, vol.~23, no.~6, pp. 6438--6452, Jun. 2023.

\bibitem{hao2022securing}
W.~Hao, J.~Li, G.~Sun, M.~Zeng, and O.~A. Dobre, ``Securing reconfigurable intelligent surface-aided cell-free networks,'' \emph{IEEE Trans. Inf. Forensics Secur.}, vol.~17, pp. 3720--3733, 2022.

\bibitem{sihlbom2022reconfigurable}
B.~Sihlbom, M.~I. Poulakis, and M.~Di~Renzo, ``Reconfigurable intelligent surfaces: Performance assessment through a system-level simulator,'' \emph{IEEE Wireless Commun.}, vol.~30, no.~4, pp. 98--106, Apr. 2022.

\bibitem{9244106}
A.~Zappone, M.~Di~Renzo, X.~Xi, and M.~Debbah, ``On the optimal number of reflecting elements for reconfigurable intelligent surfaces,'' \emph{IEEE Wireless Commun. Lett.}, vol.~10, no.~3, pp. 464--468, Mar. 2021.

\bibitem{an2021low}
J.~An, C.~Xu, L.~Gan, and L.~Hanzo, ``Low-complexity channel estimation and passive beamforming for {RIS}-assisted {MIMO} systems relying on discrete phase shifts,'' \emph{IEEE Trans. Commun.}, vol.~70, no.~2, pp. 1245--1260, Feb. 2021.

\bibitem{doi:10.1126/science.aat8084}
X.~Lin, Y.~Rivenson, N.~T. Yardimci, M.~Veli, Y.~Luo, M.~Jarrahi, and A.~Ozcan, ``All-optical machine learning using diffractive deep neural networks,'' \emph{Sci.}, vol. 361, no. 6406, pp. 1004--1008, Jul. 2018.

\bibitem{liu2022programmable}
C.~Liu, Q.~Ma, Z.~J. Luo, Q.~R. Hong, Q.~Xiao, H.~C. Zhang, L.~Miao, W.~M. Yu, Q.~Cheng, L.~Li \emph{et~al.}, ``A programmable diffractive deep neural network based on a digital-coding metasurface array,'' \emph{Nat. Electro.}, vol.~5, no.~2, pp. 113--122, Feb. 2022.

\bibitem{nerini2024physically}
M.~Nerini and B.~Clerckx, ``Physically consistent modeling of stacked intelligent metasurfaces implemented with beyond diagonal {RIS},'' \emph{IEEE Commun. Lett.}, vol.~28, no.~7, pp. 1693--1697, Jul. 2024.

\bibitem{liu2024stacked}
H.~Liu, J.~An, X.~Jia, L.~Gan, G.~K. Karagiannidis, B.~Clerckx, M.~Bennis, M.~Debbah, and T.~J. Cui, ``Stacked intelligent metasurfaces for wireless communications: Applications and challenges,'' \emph{IEEE Wireless Commun.}, vol.~32, no.~4, pp. 46--53, Apr. 2025.

\bibitem{shi2025downlink}
E.~Shi, J.~Zhang, J.~An, G.~Zhang, Z.~Liu, C.~Yuen, and B.~Ai, ``Joint ap-ue association and precoding for sim-aided cell-free massive mimo systems,'' \emph{IEEE Trans. Wireless Commun.}, vol.~24, no.~6, pp. 5352--5367, Jun. 2025.

\bibitem{10158690}
J.~An, C.~Xu, D.~W.~K. Ng, G.~C. Alexandropoulos, C.~Huang, C.~Yuen, and L.~Hanzo, ``Stacked intelligent metasurfaces for efficient holographic {MIMO} communications in {6G},'' \emph{IEEE J. Sel. Areas Commun.}, vol.~41, no.~8, pp. 2380--2396, Aug. 2023.

\bibitem{nadeem2023hybrid}
J.~An, A.~Chaaban \emph{et~al.}, ``Hybrid digital-wave domain channel estimator for stacked intelligent metasurface enabled multi-user {MISO} systems,'' \emph{2024 IEEE Wireless Communications and Networking Conference (WCNC)}, pp. 1--6, 2024.

\bibitem{li2025sim}
Z.~Li, J.~An, and C.~Yuen, ``Stacked intelligent metasurface-enhanced wideband multiuser mimo ofdm-im communications,'' \emph{arXiv preprint arXiv:2509.22327}, 2025.

\bibitem{wang2024multi}
Z.~Wang, H.~Liu, J.~Zhang, R.~Xiong, K.~Wan, X.~Qian, M.~Di~Renzo, and R.~C. Qiu, ``Multi-user {ISAC} through stacked intelligent metasurfaces: New algorithms and experiments,'' in \emph{GLOBECOM 2024-2024 IEEE Global Communications Conference}.\hskip 1em plus 0.5em minus 0.4em\relax IEEE, 2024, pp. 4442--4447.

\bibitem{liu2025stacked}
H.~Liu, J.~An, X.~Jia, L.~Gan, G.~K. Karagiannidis, B.~Clerckx, M.~Bennis, M.~Debbah, and T.~J. Cui, ``Stacked intelligent metasurfaces for wireless communications: Applications and challenges,'' \emph{IEEE Wireless Commun.}, vol.~32, no.~4, pp. 46--53, Apr. 2025.

\bibitem{pei2024stacked}
C.~Pei, K.~Huang, L.~Jin, X.~Xu, Y.~Zhou, and Y.~Guo, ``Stacked intelligent metasurfaces assisted integrated-sensing-and-resistance anti jamming,'' \emph{IEEE Commun. Lett.}, vol.~29, no.~2, pp. 383--387, Feb. 2024.

\bibitem{niuab2024enhancing}
H.~Niuab, J.~An, L.~Zhang, X.~Lei, and C.~Yuen, ``Enhancing physical layer security for {SISO} systems using stacked intelligent metasurfaces,'' in \emph{2024 IEEE VTS Asia Pacific Wireless Communications Symposium (APWCS)}.\hskip 1em plus 0.5em minus 0.4em\relax IEEE, 2024, pp. 1--5.

\bibitem{niu2024efficient}
H.~Niu, X.~Lei, J.~An, L.~Zhang, and C.~Yuen, ``On the efficient design of stacked intelligent metasurfaces for secure {SISO} transmission,'' \emph{IEEE Trans. Inf. Forensics Secur.}, vol.~20, no.~11, pp. 60--70, Nov. 2024.

\bibitem{yu2024nature}
Z.~Yu, H.~Li, W.~Zhao, P.-S. Huang, Y.-T. Lin, J.~Yao, W.~Li, Q.~Zhao, P.~C. Wu, B.~Li \emph{et~al.}, ``High-security learning-based optical encryption assisted by disordered metasurface,'' \emph{Nat. commun.}, vol.~15, no.~1, p. 2607, Jan. 2024.

\bibitem{alexandropoulos2022pervasive}
G.~C. Alexandropoulos, K.~Stylianopoulos, C.~Huang, C.~Yuen, M.~Bennis, and M.~Debbah, ``Pervasive machine learning for smart radio environments enabled by reconfigurable intelligent surfaces,'' \emph{Proc. IEEE}, vol. 110, no.~9, pp. 1494--1525, Sep. 2022.

\bibitem{zhang2025marl}
J.~Zhang, Z.~Liu, Y.~Zhu, E.~Shi, B.~Xu, C.~Yuen, D.~Niyato, M.~Debbah, S.~Jin, B.~Ai \emph{et~al.}, ``Multi-agent reinforcement learning in wireless distributed networks for {6G},'' \emph{arXiv preprint arXiv:2502.05812}, 2025.

\bibitem{zhu2024robust}
F.~Zhu, X.~Wang, C.~Huang, Z.~Yang, X.~Chen, A.~Alhammadi, Z.~Zhang, C.~Yuen, and M.~Debbah, ``Robust beamforming for {RIS}-aided communications: Gradient-based manifold meta learning,'' \emph{IEEE Trans. Wireless Commun.}, vol.~23, no.~11, pp. 15\,945--15\,956, Nov. 2024.

\bibitem{dong2024secure}
R.~Dong, B.~Wang, K.~Cao, J.~Tian, and T.~Cheng, ``Secure transmission design of {RIS} enabled {UAV} communication networks exploiting deep reinforcement learning,'' \emph{IEEE Trans. Veh. Technol.}, vol.~73, no.~6, pp. 8404--8419, Jun. 2024.

\bibitem{guo2024ris}
L.~Guo, J.~Jia, J.~Chen, S.~Yang, Y.~Xue, and X.~Wang, ``Ris-aided secure {A2G} communications with coordinated multi-{UAVs}: A hybrid {DRL} approach,'' \emph{IEEE Trans. Netw. Sci. Eng.}, vol.~11, no.~5, pp. 4536--4550, May 2024.

\bibitem{khoshafa2024ris}
M.~H. Khoshafa, O.~Maraqa, J.~M. Moualeu, S.~Aboagye, T.~M. Ngatched, M.~H. Ahmed, Y.~Gadallah, and M.~Di~Renzo, ``{RIS}-assisted physical layer security in emerging {RF} and optical wireless communication systems: A comprehensive survey,'' \emph{IEEE Commun. Surveys Tuts., early access}, 2024.

\bibitem{an2023stacked2}
J.~An, M.~Di~Renzo, M.~Debbah, H.~Vincent~Poor, and C.~Yuen, ``Stacked intelligent metasurfaces for multiuser downlink beamforming in the wave domain,'' \emph{IEEE Trans. Wireless Commun.}, vol.~24, no.~7, pp. 5525--5538, Jul. 2025.

\bibitem{11182313}
E.~Shi, J.~Zhang, J.~An, M.~D. Renzo, B.~Ai, and C.~Yuen, ``Energy-efficient {SIM}-assisted communications: How many layers do we need?'' \emph{IEEE Trans. Wireless Commun., early access}, 2025.

\bibitem{an2024stacked}
J.~An, C.~Yuen, C.~Xu, H.~Li, D.~W.~K. Ng, M.~Di~Renzo, M.~Debbah, and L.~Hanzo, ``Stacked intelligent metasurface-aided {MIMO} transceiver design,'' \emph{IEEE Wireless Commun.}, vol.~31, no.~4, pp. 123--131, Apr. 2024.

\bibitem{bjornson2020rayleigh}
E.~Bj{\"o}rnson and L.~Sanguinetti, ``Rayleigh fading modeling and channel hardening for reconfigurable intelligent surfaces,'' \emph{IEEE Wireless Commun. Lett.}, vol.~10, no.~4, pp. 830--834, Apr. 2020.

\bibitem{sanayei2004antenna}
S.~Sanayei and A.~Nosratinia, ``Antenna selection in {MIMO} systems,'' \emph{IEEE Commun. Mag.}, vol.~42, no.~10, pp. 68--73, Oct. 2004.

\bibitem{ng2014robust}
D.~W.~K. Ng, E.~S. Lo, and R.~Schober, ``Robust beamforming for secure communication in systems with wireless information and power transfer,'' \emph{IEEE Trans. Wireless Commun.}, vol.~13, no.~8, pp. 4599--4615, Aug. 2014.

\bibitem{zhu2024marl}
Y.~Zhu, E.~Shi, Z.~Liu, J.~Zhang, and B.~Ai, ``Multi-agent reinforcement learning-based joint precoding and phase shift optimization for {RIS}-aided cell-free massive {MIMO} systems,'' \emph{IEEE Trans. Veh. Technol.}, vol.~73, no.~9, pp. 14\,015--14\,020, Apr. 2024.

\bibitem{zhu2024joint}
Y.~Zhu, J.~Zhang, E.~Shi, Z.~Liu, C.~Yuen, D.~Niyato, and B.~Ai, ``Joint precoding and phase shift design for ris-aided cell-free massive mimo with heterogeneous-agent trust region policy,'' \emph{IEEE Trans. Veh. Technol.}, Jan. 2025.

\bibitem{liu2025onboard}
M.~Liu, X.~Li, J.~An, and C.~Yuen, ``Onboard terrain classification via stacked intelligent metasurface-diffractive deep neural networks from sar level-0 raw data,'' \emph{ICLR ML4RS Workshop}, 2025.

\bibitem{zhu2025joint}
Y.~Zhu, J.~Zhang, E.~Shi, Z.~Liu, C.~Yuen, and B.~Ai, ``Joint power allocation and phase shift design for stacked intelligent metasurfaces-aided cell-free massive mimo systems with marl,'' \emph{arXiv preprint arXiv:2502.19675}, 2025.

\bibitem{rappaport2015wideband}
T.~S. Rappaport, G.~R. MacCartney, M.~K. Samimi, and S.~Sun, ``Wideband millimeter-wave propagation measurements and channel models for future wireless communication system design,'' \emph{IEEE Trans. Commun.}, vol.~63, no.~9, pp. 3029--3056, Sep. 2015.

\end{thebibliography}

\end{document}